\documentclass{iopjournal}
\usepackage{lmodern}
\usepackage{macros}
\usepackage{duckuments}
\usepackage{standalone}
\usepackage{amsmath}
\usepackage{amssymb}
\usepackage{bm}
\usepackage{tikz}
\usepackage[nameinlink]{cleveref}

\DeclareMathSymbol{\sm}{\mathbin}{AMSa}{"39}

\begin{document}

\articletype{Paper}

\title{Divertor topology and vacuum vessel design for stellarators}

\author{Andrew Giuliani$^{1,\dagger}$\orcid{0000-0002-4388-2782}, Raffael Wendlinger$^2$\orcid{0009-0001-8080-2657}, Misha Padidar$^1$\orcid{0000-0002-0710-4377}, Robert Davies$^3$\orcid{0000-0001-5570-5882}, Shibabrat Naik$^5$\orcid{0000-0001-7964-2513}, Calvin Lowe$^4$\orcid{0009-0009-1975-7322}, and Georg Harrer$^4$\orcid{0000-0002-1150-3987}}

\affil{$^1$Center for Computational Mathematics, Flatiron Institute, New York, NY 10010, USA}

\affil{$^2$Institute of Applied Physics, Technische Universit\"at Wien, Vienna, Austria}

\affil{$^3$Max Planck Institute for Plasma Physics, Wendelsteinstra\ss e 1, 17491 Greifswald, Germany}

\affil{$^4$Department of Physics, Hampton University, Hampton, VA 23663, USA}

\affil{$^5$Department of Mathematics, Hampton University, Hampton, VA 23663, USA}

\affil{$^\dagger$Author to whom any correspondence should be addressed.}

\email{agiuliani@flatironinstitute.org}

\begin{abstract}
    We present stellarator optimization algorithms for designing the edge magnetic structure in vacuum fields, together with the vacuum vessel.
    First, we introduce a numerical method that robustly computes periodic-orbit fixed points of any type (elliptic, hyperbolic, or parabolic), which could form the basis of a divertor.  
    To couple divertor and vacuum vessel design, we introduce parametric families of vacuum vessels for which point-to-vessel distances, and their derivatives, can be computed efficiently.
    The resulting algorithms use signed distance functions to enforce coil-vessel clearance while allowing coils to be placed on or off the vessel.
    Using these methods, we jointly optimize modular coils and the vacuum vessel to realize a wide range of magnetic topologies for diverting exhaust, including standard X-point divertors, and single- and double-null configurations. For the first time, we show that precise snowflake divertors can be achieved in stellarators.
    Using this framework, we generate a number of quasi-axisymmetric stellarator designs with compatible vacuum vessels and diverse divertor architectures, which we consider to be candidates for a next-generation STAR\_Lite prototype.
\end{abstract}

\section{Introduction}
\label{sec:intro}

The divertor is an essential mechanism in tokamaks and stellarators for safely exhausting heat and particles. By routing escaping plasma towards reinforced divertor plates, the heat load on plasma-facing components can be kept within engineering limits, and helium ``ash" and other impurities (sputtered wall material, for example) can be steadily removed by nearby pumps \cite{stangeby2000tutorial, konig2002divertor, loarte2001effects, boozer2015stellarator, gates2017recent, krieger2025scrape}. In addition, tokamak experiments have emphasized that the efficiency of diversion can also have significant consequences for the plasma core, for example: proper diversion can lead to steep profile gradients, and the ``high-performance mode" \cite{akers2002h, akers2003transport, ryutov_snowflake_2015}, and  small resonant magnetic perturbations in the edge can mitigate potentially destructive ``edge-localized modes", in which a large fraction of plasma stored energy is rapidly expelled \cite{evans2015resonant, yang2024tailoring}. In stellarators, the Large Helical Device (LHD) has also reported a synergy between divertor and core performance, with ``closed divertors" leading to an improvement in core plasma density, the establishment of an internal transport barrier and reduced impurity retention in the core \cite{ohyabu2006observation, yamada2011overview}. 
Incorporating divertor design as early in the design cycle as possible is therefore likely to improve the commercial viability of fusion power plants. 

Following the success of Wendelstein 7-X \cite{renner2004, grulke2024overview, grulke2026overview}, the island divertor has emerged as a common choice in stellarator design, featured in numerous recent designs \cite{Warmer_2022, lion2025stellaris, hegna2025infinity, bader2025power, goodman_squid_2025, davies_squid_2025, sanchez2026ciemat, lazerson2026fixed}.
Recently, however, there have been efforts to design tokamak-style X-point divertors (i.e. diverting X-points that sit at the top and/or bottom of the device and therefore have a rotational transform $\iota=0$), to leverage the wealth of knowledge and experience gained from 50 years of research in tokamaks \cite{harrer_star_lite_2026,gates_stellarator_2025}. For island, helical, or $\iota=0$ X-point divertors, the structure of the edge magnetic field (in particular, the periodic magnetic field lines, also known as ``fixed points"; X-points and O-points) is central to the divertor behavior. Despite the importance of the divertor in stellarators, the vast range of divertor possibilities has been explored relatively little, and divertor-related design efforts and optimization tool development (arguably) lag progress compared to methods for optimizing the plasma core.  

Controlling the geometry of the tokamak divertor is, relatively speaking, a much more mature field of research \cite{forge}, and there is a taxonomy of proposed (and experimentally realized) exhaust diversion structures. The creation of either one X-point (``single-null") or an X-point at the top and bottom of the device (``double-null") is long established (see e.g. \cite{stangeby1990plasma, stangeby2000tutorial}) and has been incorporated into the design of the large tokamaks (for example ITER \cite{aymar2002iter}, JET \cite{keilhacker2001scientific}, JT-60 \cite{yoshikawa1987overview, di2014overview}, ASDEX \cite{keilhacker1985asdex}, SPARC \cite{kuang2020divertor, rodriguez2022overview}). More advanced tokamak divertor experiments have been performed, for example by tightly baffling the divertor (i.e. preventing the neutralized particles from re-entering the plasma via the plasma-facing component geometry) \cite{kuang2020divertor, rodriguez2022overview, sun2023performance}. A second example is ``long-legged" or ``Super-X" divertors, in which the divertor leg is extended to a region of larger major radius and weaker magnetic field, thus increasing the area of PFCs wetted by the plasma \cite{valanju2009super, gallo2018impact, harrison2024benefits}. 
Yet another promising avenue is the ``snowflake" divertor, in which two X-points are ``brought together" to form a second-order null in the poloidal magnetic field, resulting in a six-legged fixed point \cite{ryutov2007geometrical, ryutov2008magnetic, ryutov2012snowflake}. These have been shown to be efficient at spreading due to increased number of legs, increased upstream-to-target connection length and enhanced turbulent transport around the fixed point (``churning modes") \cite{ryutov2007geometrical,ryutov2008magnetic, soukhanovskii_developing_2018, soukhanovskii2022first,  gorno2023power, ryutov2014churning, power2025simulations}. Third-order poloidal null (``cloverleaf") tokamak configurations have also been proposed \cite{ryutov2013divertor, kuang2026implementation}. The novel and wide-ranging divertor options in a tokamak are useful guides, because all can (in principle) be attained in a stellarator, the latter having greater freedom via the breaking of axisymmetry. 

\begin{figure}[tbh!]
    \centering
    \includegraphics[width=\linewidth]{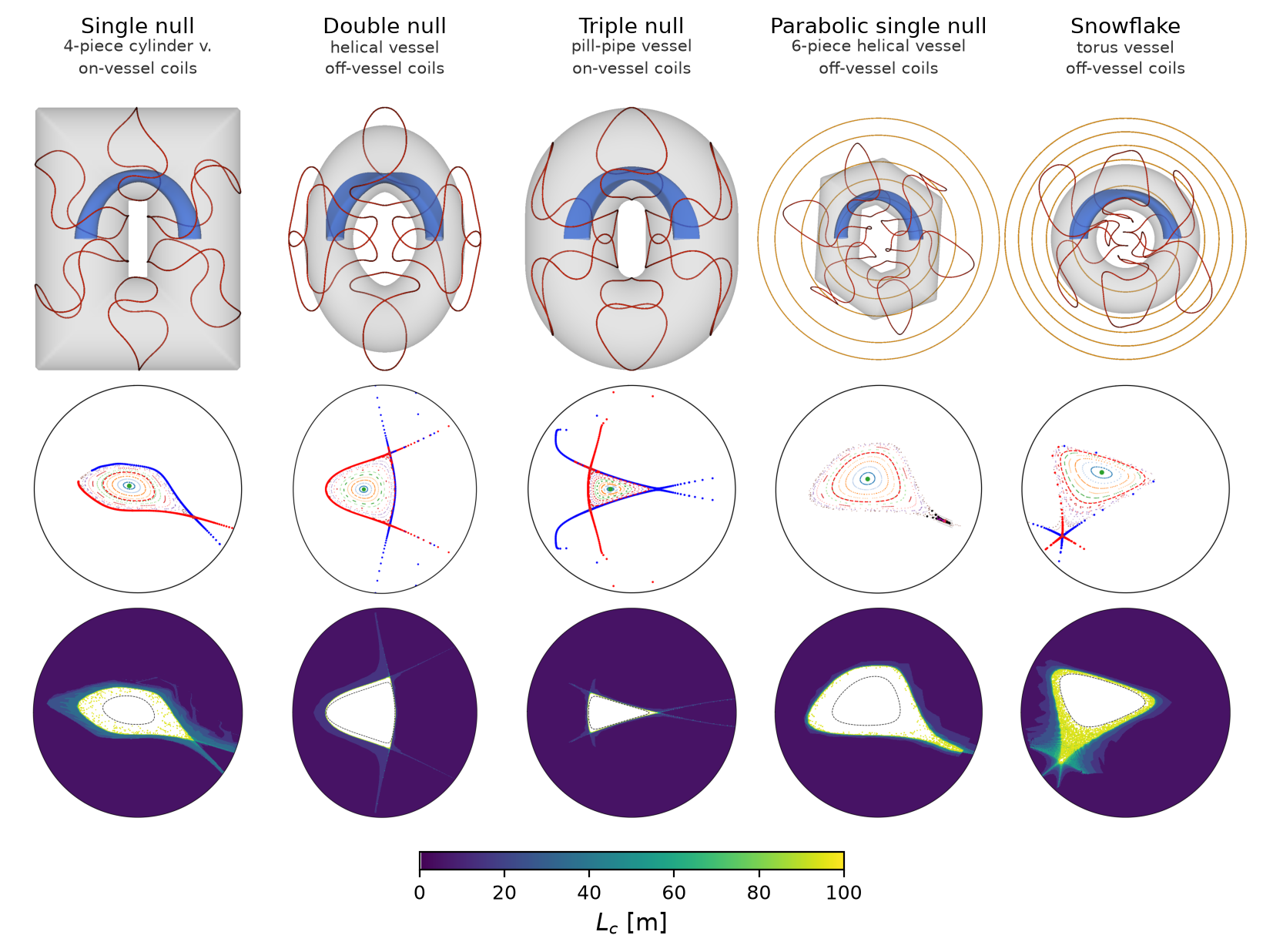}
    \caption{STAR\_Lite class stellarators featuring various kinds of divertors, designed using the algorithms in this work. 
    The top row shows the magnetic surface on a field period, the six modular coils with the vacuum vessel geometry (gray).
    The middle row shows Poincar\'e sections of the devices in the $\phi=0$ plane, with stable and unstable manifolds in red and blue.
    The bottom row depicts the connection length $L_c$ at $\phi=0$ of each device.
    The black circle constitutes the vacuum vessel.
    }
    \label{fig:divertor_zoo}
\end{figure}

This work introduces a method for computing fixed points of a vacuum magnetic field and controlling their positions and type within a stellarator optimization loop. The method is numerically stable when computing fixed points of any type, and admits derivatives for use in stellarator optimization. To ensure compatibility of the fixed points with a vacuum vessel, we treat the vacuum vessel geometry as degrees of freedom, in addition to the coil geometries. To efficiently compute distances from the fixed points to the vessel, we introduce families of vacuum vessel geometries for which distances from points to the vessel have closed-form expressions, or only require solving a one-dimensional root finding problem. 
Through numerical experiments, we demonstrate how to optimize for nested flux surface geometry, quasisymmetry, magnetic well, and the position and even the type (elliptic, hyperbolic, parabolic) of divertor fixed point. We showcase a set of single and double null, hyperbolic and parabolic divertors, and for the first time, we show that non-axisymmetric snowflake divertors (a special subset of six-legged parabolic divertors) can be achieved in a stellarator. 
\Cref{fig:divertor_zoo} shows five devices designed with the methods in this paper, including single-null, double-null, and triple-null X-point divertors, a parabolic single-null divertor, and a snowflake.

An indirect method of ``optimizing" the edge for an island divertor is equilibrium (``stage one") optimization, by controlling the $\iota$ profile within the confined region, since magnetic islands depend on a low-order rational $\iota$ and the magnetic shear at this location \cite{sanchez2026ciemat, goodman_squid_2025, bader2025power, hegna2025infinity, lazerson2026fixed}.  Such an approach can be used to approximately fix the location and width of the magnetic island chain, but cannot (for example) control the phase of the island chain, i.e., the poloidal location of the magnetic islands.
In addition, the island size and degree of magnetic chaos also depend upon the radial magnetic field (i.e. resonant perturbations) with respect to a set of hypothetical nested surfaces in the edge, which depends on the coils rather than the magnetic equilibrium. Another way to optimize island size is by targeting Greene's residue for X- and O-points in the island; for example, shrinking islands in the confined region by minimizing the residue \cite{Geraldini_Landreman_Paul_2021}. On the experimental side, the island size in W7-X is routinely adjusted using windowpane-shaped magnetic coils \cite{andreeva2022magnetic, davies2026characterisation}, and more recently exotic edge configurations (``giant islands") have been proposed in W7-X and HSX, based on empirical simulation results \cite{davies2026computationalstudiesgiantedge}. Non-island divertor edge optimization includes controlling stellarator equilibria to have high-curvature on the boundary \cite{gaur2026omnigenous} or targeting fixed points directly using a $\mathbf B \times d\mathbf l$ minimization \cite{FLOM2026115846}.

The remainder of the paper is organized as follows. In \cref{sec:ODE} we introduce a well-conditioned spectral method for computing periodic field lines and their fixed points, regardless of type (elliptic, hyperbolic, or parabolic). \Cref{sec:vessel} presents a family of vacuum vessel geometries whose signed distance functions can be evaluated rapidly. In \cref{sec:optimization} we combine these ingredients into a single framework that jointly optimizes the modular coils, divertor fixed points, and vacuum vessel. \Cref{sec:example_configs} presents representative configurations.

\section{Periodic field lines}
\label{sec:field_lines}
Periodic field lines of the magnetic field, beyond the region of nested flux surfaces, are the basis of the magnetic structures that enable diversion of ash particles to divertor plates. In this section, we review how these field lines are computed as fixed points of a Poincar\'e return map, and how the trace of the return map's Jacobian classifies each fixed point as hyperbolic, elliptic, or parabolic, resulting in distinct divertor structures. The mathematics introduced here sets the stage for the numerical methods described in \cref{sec:spectral_method} for computing fixed points.

\subsection{Field line equations and the tangent map} 
A closed field line, parametrized to have uniform incremental arclength, satisfies the ordinary differential equation (ODE)
\begin{equation}
\begin{aligned} \label{eq:fieldline}
\mathbf r_f(t) := \frac{\bm{\Gamma}'(t)}{L} - \frac{\mathbf B(\bm{\Gamma}(t))}{\|\mathbf B(\bm{\Gamma}(t))\|}=0,\\
\bm\Gamma(0) = \bm \Gamma_0
\end{aligned}
\end{equation}
where $\bm{\Gamma}(t) = [x(t), y(t), z(t)]$ is the Cartesian coordinates of the field line, $L$ is the (constant) length of the field line from its start until it closes on itself, $\bm{\Gamma}_0=[\Gamma_{0,x}, \Gamma_{0,y}, \Gamma_{0,z}]$ is the initial field line position, and $t \in [0, t_p)$ parameterizes the curve, where $t_p = 1/\nfp$ is the period. While it is common to express \cref{eq:fieldline} in cylindrical coordinates $(R(\phi), \phi,Z(\phi))$, we opt for the more general Cartesian representation, which affords us more flexibility when handling strongly shaped devices and their field lines.
Computing periodic field lines can be cast as searching for roots of the displacement map,
\begin{equation}\label{eq:displacement}
\begin{aligned}
    \bm D(\bm\Gamma_0, L) :&= \bm \Gamma(t_p; \bm \Gamma_0, L) - \bm{\mathcal R}(t_p)\bm \Gamma_0=0, \\
    \Gamma_{0,y} &= 0,
\end{aligned}
\end{equation}
where $\bm \Gamma(t)$ is governed by \eqref{eq:fieldline}, $\bm{\mathcal{R}}(t)$ is the standard rotation matrix about the $Z$-axis by the angle $2\pi t$, and the field line is launched from the $XZ$-plane.

Differentiating \cref{eq:fieldline} with respect to the initial position of a fixed point, $\bm \Gamma(0)$, gives the forward sensitivity equation:
\begin{equation}\label{eq:tangentmap}
\begin{aligned}
       \frac{1}{L} \bm \Phi'(t) -  \left(\bm I - \frac{\mathbf B}{\|\mathbf B\|}\frac{\mathbf B^T}{\|\mathbf B\|} \right)\frac{\nabla \mathbf B}{\|\mathbf B\|} \bm \Phi(t) &=0, \\
       \bm \Phi(0) &= \bm I.
\end{aligned}
\end{equation}
The solution $\bm \Phi$ after one period, $\bm \Phi(t_p)=\partial \bm \Gamma(t_p)/\partial \bm \Gamma(0)\in \mathbb R^{3\times 3}$, is known as the ``tangent map".
Although $\bm \Gamma(t)$ is periodic, $\bm \Phi(t)$ may not be, since field lines near the fixed point are not periodic in general.
Numerically, this implies that \cref{eq:tangentmap} should be solved with a Chebyshev collocation method, rather than a Fourier collocation method. 
Projecting the tangent map onto the plane orthogonal to the fixed point, spanned by the normal and binormal Frenet-Serret vectors $\mathbf n, \mathbf b$, gives the tangent map $\bm M\in \mathbb R^{2\times 2}$
\begin{equation}\label{eq:M_NB}
    \bm M = [\mathbf n(t_p), \mathbf b(t_p)]^T \bm \Phi(t_p)[\mathbf n(0), \mathbf b(0)].
\end{equation}
$\bm M$ is the Jacobian of the return map, and can be used to construct an approximate Poincar\'e section in the plane orthogonal to $\bm \Gamma'(0)$, in the neighborhood of the fixed point at $t=0$. The position of a field line after $t_p$ when launched from $\bm \Gamma(0) + [\mathbf n(0), \mathbf b(0)] \bm\delta_0$, for small $\bm\delta_0 \in \mathbb R^2$, is approximated by $\bm \Gamma(kt_p) + [\mathbf n(kt_p), \mathbf b(kt_p)] \bm\delta_k$, where $\bm\delta_k$ follows the linear recurrence, 
\begin{align}
    \bm\delta_{k+1} = \bm M\bm\delta_k.
    \label{eq:recurrence}
\end{align}
As a result of \cref{eq:recurrence}, field line dynamics can be understood by analyzing the tangent map $\bm M$.
As we will discuss in \Cref{sec:types_of_fixed_points}, the trace of $\bm M$ is a useful classifier of different ``flavors" of periodic orbits that can be used to design divertors.

\subsection{Types of diverting fixed points}
\label{sec:types_of_fixed_points}

The Hamiltonian nature of the field line equations guarantees $\det(\bm M)=1$, by the Abel-Liouville theorem. The eigenvalues of $\bm M$ are thus only determined by the trace. We now describe a classification of the different types of fixed points based on $\Trace(\bm M)$. Each class constitutes a different type of divertor, such as an island or an X-point.

Fixed points with $|\Trace(\bm M)|>2$ are known as hyperbolic. Also known as X-points, hyperbolic fixed points are the basis of tokamak-like divertors (\cref{fig:divertor_zoo}: columns 1-3). Particles nearby the X-point are diverted along its ``legs": eigenvectors of $\bm M$ along which nearby field lines are attracted to or repelled from the fixed point. The eigenvalues of $\bm M$ give the exponential rate of attraction or repulsion near the fixed point. 

Island divertors route exhaust through island chains whose O-points are elliptic fixed points -- $|\Trace(\bm M)|<2$.  At an elliptic fixed point, $\bm M$ is similar to a rotation matrix, i.e. has eigenvalues $\lambda_{\pm} = e^{\pm i\alpha}$, and repeated iterates of the mapping trace an ellipse.
As a result, $\bm M$ is similar to a rotation matrix with rotation angle $\alpha$,
$$
\bm M = \bm S^{\sm 1}\begin{pmatrix}
    \cos(\alpha) & \sm \sin(\alpha) \\
    \sin(\alpha) & \cos(\alpha)
\end{pmatrix}\bm S.
$$
The rotational transform, or pitch of field lines at the fixed point, is given by the formula \cite{greenesresidue}
\begin{equation} \label{eq:iota_a}
\iota = \frac{\nfp}{2\pi}\alpha.
\end{equation}

\begin{figure}
     \centering
    \makebox[\textwidth][c]{%
         \includegraphics[width=0.8\linewidth]{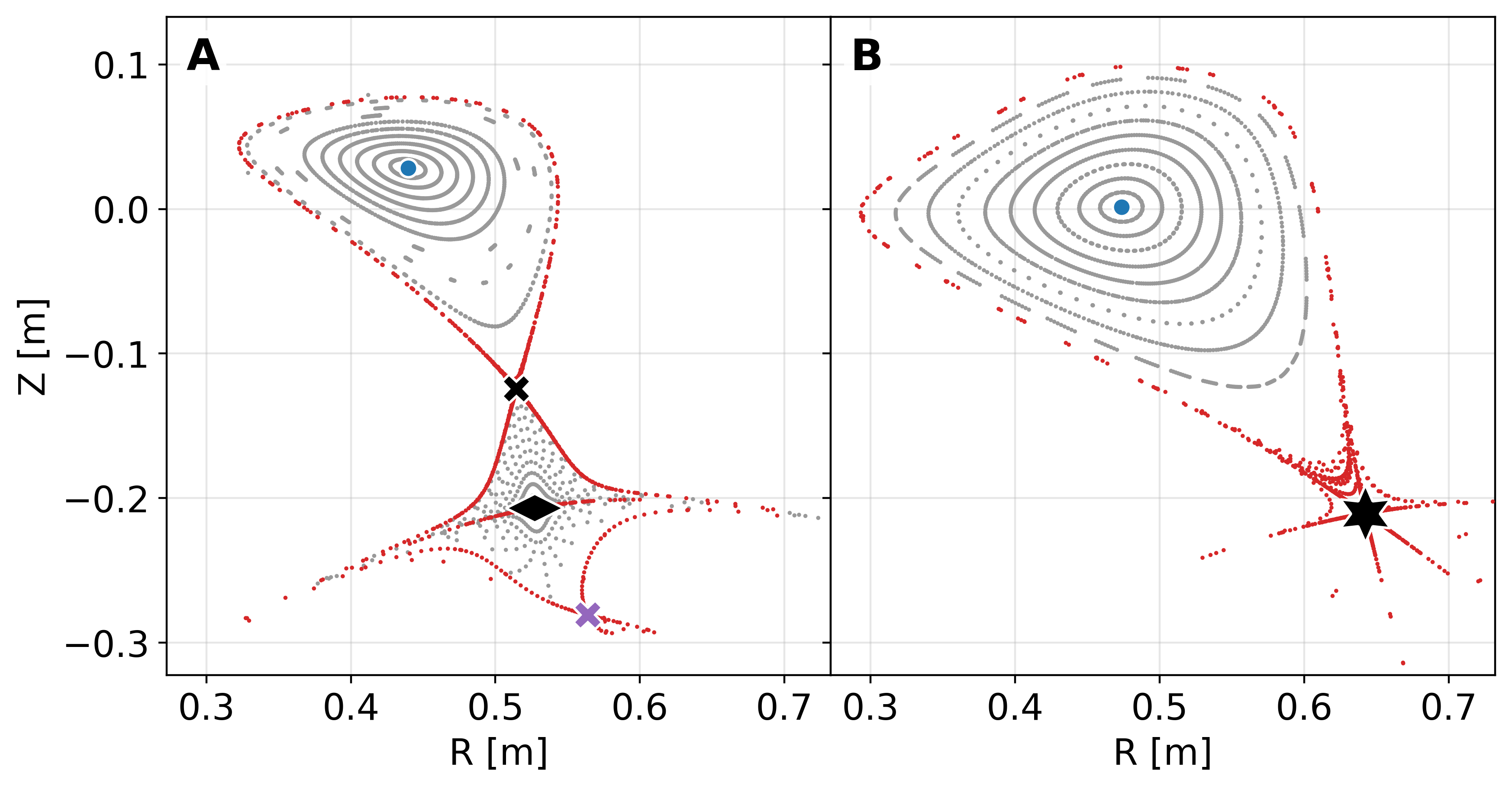}%
    }
     \caption{Stellarators exhibiting $\text{rank}(\bm M-\bm I)=0$ parabolic fixed points. Panel A shows a two-legged fixed point (horizontal diamond) with two orbiting X-points (crosses), and panel B shows a six-legged fixed point (star).}
     \label{fig:snowflake2_snowflake6}
 \end{figure}

Parabolic fixed points lie on the boundary between the elliptic ($|\Trace(\bm M)|<2$) and hyperbolic ($|\Trace(\bm M)|>2$) regimes, with $|\Trace(\bm M)| = 2$. Parabolic fixed points may not be isolated; for example, every point on a rational-$\iota$ surface is a parabolic fixed point of the return map. 
When used as part of a divertor, the fixed point is designed to be isolated. 
In this work, we give special attention to fixed points with $\Trace(\bm M) = +2$, and further classify them by the rank of the linearized displacement map $\bm M-\bm I$: \textit{rank-1 parabolic fixed points} have $\bm M\neq \bm I$ (\Cref{fig:divertor_zoo}: column 4) and \textit{rank-0 parabolic fixed points} have $\bm M=\bm I$. Rank-0 parabolic fixed points can have two or six legs protruding from the fixed point (\Cref{fig:snowflake2_snowflake6}A, \Cref{fig:divertor_zoo}: column 5).  
Those with six legs are known as \textit{snowflakes}, as shown in \Cref{fig:snowflake2_snowflake6}B.
Parabolic fixed points are also known as degenerate roots of the displacement map \eqref{eq:displacement}, since the Jacobian at the root is not full rank.
At an isolated fixed point, 
the tangent map, $\bm M$, is insufficient for characterizing the manifolds of the nonlinear map in the neighborhood of the fixed point.
We will not explore reflection-hyperbolic ($\Trace(\bm M)<\sm 2$ \cite{Davies_2025}) or negative parabolic ($\Trace(\bm M)=\sm 2$) fixed points in this work.

\section{A spectral method for computing fixed points} \label{sec:ODE}
\label{sec:spectral_method}

We now describe methods for numerically computing fixed points of any type: hyperbolic, elliptic, and parabolic. The computations rely on treating \cref{eq:fieldline} as a nonlinear root-finding problem. When the fixed point is elliptic or hyperbolic, the Jacobian of the root-finding problem is well-conditioned and can be solved with Newton's method. Systems with parabolic fixed points have a singular Jacobian; to apply Newton's method, we introduce additional equations that stabilize the system. We begin by discussing a method for computing hyperbolic and elliptic fixed points. In both cases, the solution of the fixed point system can be differentiated with respect to system parameters using an adjoint method, making it amenable to gradient-based stellarator optimization.

\subsection{Computing hyperbolic and elliptic fixed points}
In this section, we outline a numerical method for computing elliptic and hyperbolic fixed points.  
This is done by solving \eqref{eq:fieldline} using a spectral collocation method with Newton's method. 

The fixed point curve is parameterized using the \texttt{CurveXYZFourierSymmetries} representation in \texttt{SIMSOPT} \cite{landreman_simsopt_2021}, originally proposed in \cite{10.1063/5.0226688}. This curve parametrization can efficiently represent stellarator symmetric and $\nfp$-periodic field lines. The parametrization expresses $\bm \Gamma(t) = [x(t), y(t), z(t)]$, $t\in[0,1)$ as:
\begin{equation}
\begin{aligned}\label{eq:curvxyzfouriersymmetries1}
    \hat x(t) &= \hat x_{c, 0} +  \sum_{m=1}^{N_F} \hat x_{c,m} \cos(2\pi m \nfp t) +\hat x_{s,m} \sin(2\pi m \nfp t) \\
    \hat y(t) &= \hat y_{c, 0} +  \sum_{m=1}^{N_F} \hat y_{c,m} \cos(2\pi m\nfp t) +\hat y_{s,m} \sin(2\pi m\nfp t) \\
    z(t) &= z_{c, 0} +  \sum_{m=1}^{N_F} z_{c,m} \cos(2\pi m\nfp t) +z_{s,m} \sin(2\pi m\nfp t), \\
\end{aligned}
\end{equation}
where the coordinates for $x(t), y(t)$ are:
\begin{equation}
\begin{bmatrix}
x(t) \\
y(t)\\
\end{bmatrix} = \bm{\mathcal{R}}(t) \begin{bmatrix}
\hat x(t) \\
\hat y(t)
\end{bmatrix}. \label{eq:curvxyzfouriersymmetries2}
\end{equation}
For non-stellarator symmetric curves, this curve parametrization has $3(2N_F+1)$ Fourier harmonics, also called degrees of freedom for the optimization.
For stellarator symmetric curves, the symmetry breaking harmonics are omitted from \eqref{eq:fieldline}, i.e., $\hat{x}(t)$ only has cosine harmonics, while $\hat{y}(t)$, $z(t)$ only have sine harmonics, resulting in $3N_F+1$ degrees of freedom.

To apply the spectral collocation method, \cref{eq:fieldline} is discretized by evaluating the residuals at collocation points, $t_k = k\,t_p/(2N_F+1)$ for $k = 0, 1, \hdots, 2N_F$,
\begin{equation}\label{eq:fieldline_2}
    \mathbf r_f(t_k; \mathbf x) = 0 \text{ for } k =0, 1, \hdots, 2N_F,
\end{equation}
where $\mathbf x$ are the Fourier harmonics in \eqref{eq:curvxyzfouriersymmetries1}, \eqref{eq:curvxyzfouriersymmetries2}, and the curve position, $\bm\Gamma(t)$, is evaluated using the parametrization \eqref{eq:curvxyzfouriersymmetries2}.

When the magnetic field is not-stellarator symmetric, the curve is not fully specified by this system of residuals, so we introduce an additional constraint on the field line: $y(0) = 0$, resulting in $3(2N_F+1)+1$ equations. 
When the magnetic field is stellarator symmetric, $y(0) = 0$ is always satisfied by the parametrization, thus this additional equation is not added to the system. In addition, the $X$-component of the residual for $t=0$ is always zero: $r_x(0)=0$, regardless of the curve harmonics, because for a stellarator symmetric curve and stellarator symmetric field, $B_x(\bm \Gamma(0))=0$ implies $x'(0) = 0$. Lastly, half of the remaining residual equations are redundant due to stellarator symmetry, since $\mathbf r_f(\sm t_k) = [r_{f,x}(\sm t_k), r_{f,y}(\sm t_k), r_{f,z}(\sm t_k)] = [\sm r_{f,x}(t_k), r_{f,y}(t_k), r_{f,z}(t_k)]$. Removing these unnecessary equations, we are left with only $3N_F+2$ independent residuals.
Considering $L$ to be unknown in \eqref{eq:fieldline}, the total number of unknowns is $3(2N_F+1)+1$ and $3N_F+2$ in the non-stellarator symmetric and stellarator symmetric cases, respectively.
In both cases, \eqref{eq:fieldline_2} along with the mentioned modifications result in a balanced system of nonlinear equations and unknowns, which we solve using Newton's method.

\subsection{Numerically approximating parabolic fixed points}\label{sec:parabolic}

When the fixed point is parabolic, the Jacobian of the system \cref{eq:fieldline_2} can be singular or poorly conditioned.
When applied to systems with singular Jacobians, Newton's method can converge with only linear accuracy \cite{suli2003introduction}, and may be prone to diverging \cite{griewank1980starlike}.
Moreover, using the fixed point solver within a stellarator optimization loop requires discretely exact derivatives of the fixed point position with respect to the design variables, which are challenging to accurately compute when the fixed point is degenerate. To address these shortcomings, we augment the system of equations with additional constraints that increase the rank of the Jacobian.

We illustrate the source of this ill-conditioning with a simple example. Consider searching for a fixed point at cylindrical $\phi=0$ in the $(R,Z)$ plane by solving for $(R_0, Z_0)$ such that
\begin{equation}\label{eq:return}
\mathbf D(R_0, Z_0)=
\begin{pmatrix}
    R(t_p, R_0, Z_0) - R_0  \\
    Z(t_p, R_0, Z_0) - Z_0 
\end{pmatrix}=0,
\end{equation}
where $R(t_p, R_0, Z_0), Z(t_p, R_0, Z_0)$ denote the position of the field line launched from $(R_0, Z_0)$ after one field period. Newton's method relies on the Jacobian of \eqref{eq:return}, given by 
\begin{equation} \label{eq:jacobian}
\mathbf  J(R_0, Z_0) = \begin{pmatrix}
    \partial R/\partial R_0 & \partial R/\partial Z_0 \\
    \partial Z/\partial R_0 & \partial Z/\partial Z_0
\end{pmatrix}-\bm I = \bm M-\bm I.
\end{equation}
The first term in the Jacobian is the tangent map $\bm M$ at the solution in the $RZ$-plane; the $\mathbf n, \mathbf b$ representation is given in \cref{eq:M_NB} and the two representations of $\bm M$ have the same trace and determinant.
Given that $\det(\bm M)=1$ and defining $T = \Trace(\bm M)$, the eigenvalues of \cref{eq:jacobian} are $T/2-1\pm\sqrt{T^2-4}/2$, i.e., two zero eigenvalues when $T = 2$.

Degenerate roots of nonlinear equations can still be computed with Newton's method. To make the system well conditioned, we augment the base system of equations with additional constraints and degrees of freedom which stabilize the system \cite{griewank_reddien_1984}. 
We use two, more general, formulations to compute fixed points when the Jacobian of \eqref{eq:fieldline_2} is ill-conditioned or even singular in the neighborhood of parabolic fixed points. Both formulations require that some degrees of freedom, which can shape the magnetic field near the fixed point, be reserved as dependent degrees of freedom. The dependent degrees of freedom are solved for, jointly with the position of the fixed point, to ensure a fixed point of the desired type exists. In all numerical experiments, we introduce auxiliary poloidal field (PF) coils. The currents and positions of the PF coils are selected to ensure the fixed point exists.

The first formulation for stabilizing the fixed point equations is capable of computing parabolic fixed points, aside from those with $\bm M = \bm I$, i.e. snowflakes. While this formulation is slightly more limited than the second, it offers simplicity. The second formulation is completely general and can be used to compute fixed points of all types. In the first formulation, the standard fixed point equations,
\cref{eq:fieldline_2}, are augmented with the trace condition,
\begin{align} 
    \Trace(\bm M(\bm \eta)) &=T, \label{eq:stabilization_first}
\end{align}
where $T =2$ when considering parabolic fixed points. The vector $\bm \eta$ represents additional degrees of freedom for shaping the magnetic field and balancing the system of equations, which we will discuss momentarily. As noted earlier, the additional condition \cref{eq:stabilization_first} does not stabilize the system when $\bm M  = \bm I$. In this case, the derivative of the trace condition \cref{eq:stabilization_first} with respect to the axis position is zero; a result of $\det(\bm M)=1$. Due to this restriction, we introduce the more general, second formulation.
The second formulation augments the base discretization, \cref{eq:fieldline_2}, with four equations,
\begin{align} %
    M_{i,j}(\bm\eta) &= m_{i,j}, ~i,j =1, 2
    \label{eq:stabilization_second}
\end{align}
where the four values $M_{1,1}, M_{1,2}, M_{2,1}, M_{2,2}$ are the entries of $\bm M$, and $m_{1,1}, m_{1,2}$, $m_{2,1}, m_{2,2}$ are their associated target values. 
In practice, we do not need to constrain the fourth entry of $\bm M$ since it is fixed by $\det(\bm M)=1$. If the field line is stellarator symmetric, then $M_{1,1}=M_{2,2}$ so we do not need to constrain one of the diagonal entries.

The degrees of freedom $\bm \eta$ are introduced to build a balanced system and to introduce enough flexibility in shaping the magnetic field that a fixed point can be found. In the first formulation, only one additional degree of freedom is necessary to balance the one new equation, \cref{eq:stabilization_first}, and in the second formulation, only two additional degrees of freedom are needed. To achieve sufficient flexibility in shaping the magnetic field, it is recommended to add even more degrees of freedom, such that the system is underdetermined. In practice, we add one or more PF coils, and let $\bm \eta$ represent their current, radii, and vertical position. It may be preferable to instead use toroidal field coils, windowpane coils, or some of the modular coils as degrees of freedom. 
Although the Jacobian of underdetermined systems has a null-space, one can still take a Newton step using the pseudo-inverse of the Jacobian without sacrificing performance.

With the additional equations and degrees of freedom, Newton's method can now be applied to the joint system, \cref{eq:fieldline_2} with either \cref{eq:stabilization_first} or \cref{eq:stabilization_second}. In \Cref{fig:cond1}F, we show that the condition number of the Jacobian stays finite and bounded during the solve, unlike when solving \cref{eq:fieldline_2} directly. \Cref{fig:cond2}D shows an example of using the second formulation to manipulate a snowflake while maintaining a well-conditioned nonlinear solve.

\section{Examples of computing parabolic fixed points}
\label{sec:examples}
In this section, we walk through two numerical examples to showcase the capability of the methods introduced in \Cref{sec:spectral_method}, and to demonstrate that they are numerically stable. Starting from an initial stellarator, we perform a continuation on the trace of a fixed point through the marginal parabolic limit, pushing the fixed points through bifurcation, converting it from one type to another. In \cref{sec:rank_1_example}, O-points are converted to X-points, and in \cref{sec:rank_0_example} a broken snowflake is healed, then converted to an elliptic fixed point. This conversion capability is useful to the STAR\_Lite experiment so that various fixed points or fixed point clusters might be studied experimentally, or imperfect snowflakes might be polished by slight modifications of the PF coil currents.

\subsection{Computing rank-1 parabolic fixed points}
\label{sec:rank_1_example}

In this section, we use formulation one, introduced in \cref{sec:spectral_method}, to compute almost-parabolic and exactly parabolic rank-1 fixed points.
We track a fixed point over the course of a continuation, converting it from an elliptic fixed point to a parabolic, then to a hyperbolic, all by modifying the currents in auxiliary PF coils.
\begin{figure}[tbh!]
    \centering
    \includegraphics[width=\textwidth]{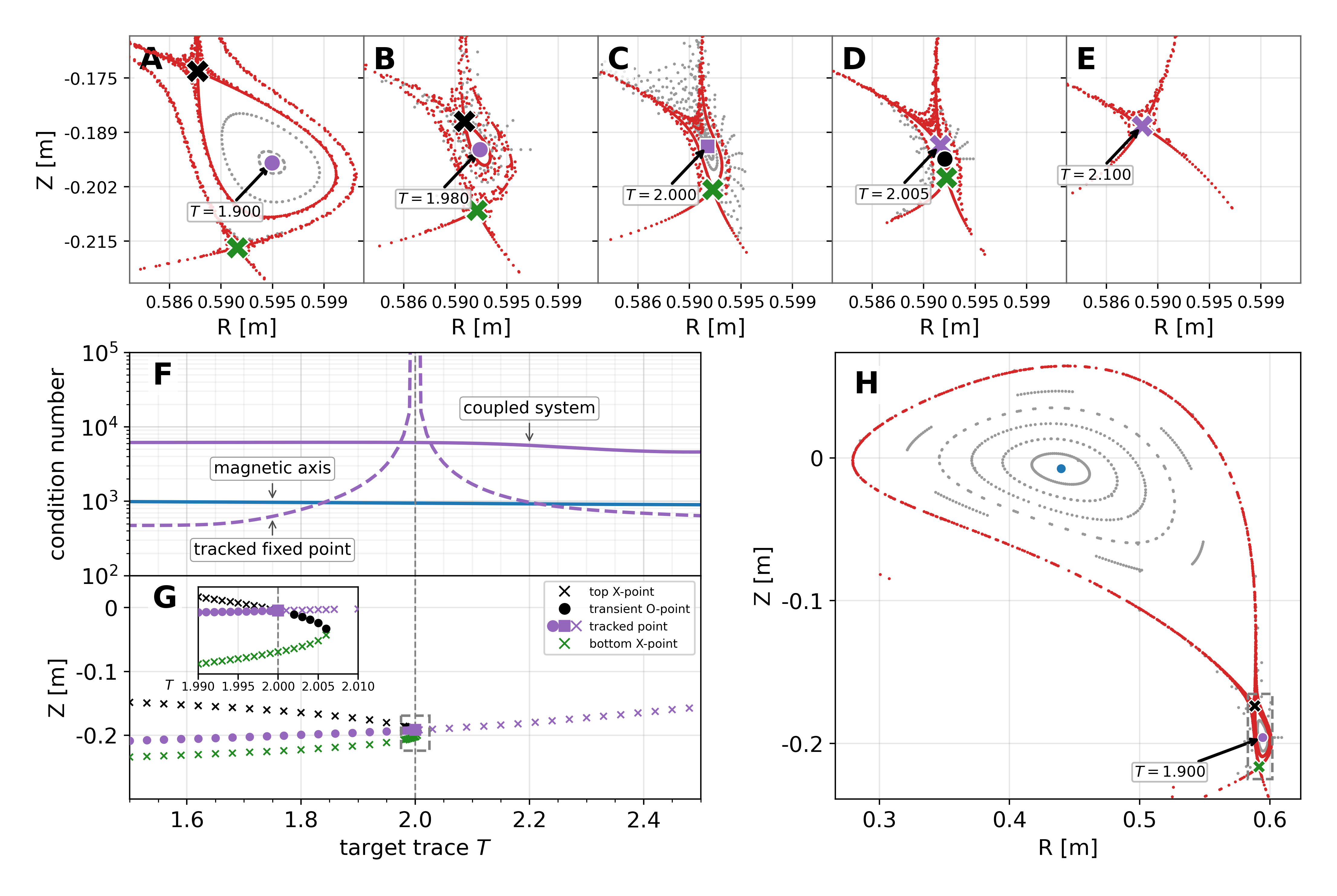}
    \caption{Condition numbers of the Jacobian of \eqref{eq:continuation-rf-a} associated to the magnetic axis (blue), of \eqref{eq:continuation-rf-o} associated to the tracked fixed point (purple dashed), and of the full system \eqref{eq:continuation} (purple solid) as the target trace $T$ of the tracked fixed point is varied.
    Circle, cross, square markers correspond respectively to O, X, and parabolic fixed points.
    Panel B contains the Z-coordinate of the fixed points illustrating the bifurcation.
    Panel C contains a Poincar\'e section, showing the full, non-stellarator symmetric device for when the trace $T=1.9$.
    Poincar\'e sections of the boxed region in \cref{fig:cond1}C at the target trace value of the tracked fixed point is varied about the bifurcation point.
    }
    \label{fig:cond1}
\end{figure}

The example uses a stellarator with modular coils and auxiliary PF coils. The modular coil currents and geometries are fixed throughout the example and are only used to generate a suitable background magnetic field to be modified by auxiliary PF coils. The PF coil geometries are also fixed, while the PF coil currents $\mathbf I_{\text{PF}}$ are varied to generate the desired fixed point.
At each step of the continuation, we solve
\begin{subequations}\label{eq:continuation}
\begin{align}
    \mathbf r_f(\mathbf x, \mathbf I_{\text{PF}}) &= 0,\label{eq:continuation-rf-a}\\
    \mathbf r_f(\mathbf t, \mathbf I_{\text{PF}}) &= 0,\label{eq:continuation-rf-o}\\
    \Trace\!\left(\bm M(\mathbf x, \mathbf I_{\text{PF}})\right) &= 1.6,\label{eq:continuation-tr-a} \\
    \Trace\!\left(\bm M(\mathbf t, \mathbf I_{\text{PF}})\right) &= T\label{eq:continuation-tr-o},
\end{align}
\end{subequations}
where $\mathbf x$ and $\mathbf t$ are degrees of freedom associated to the magnetic axis and the tracked fixed point, respectively. $\mathbf x, \mathbf t, \mathbf I_{\text{PF}}$ are varied to solve the system. The continuation varies the trace of the tangent map, $T$, from its original $T = 1.5$ to $2.0$, all the way up to $2.5$ (\cref{fig:cond1}), while maintaining the trace of the magnetic axis at $1.6$ to prevent destroying the volume of nested flux surfaces.
The system, \cref{eq:continuation}, has $3(2N_F+1)+1$ field line constraints and two additional trace conditions. 
It also has $3(2N_F+1)+1$ unknowns defining the field line and $N_{\text{PF}}$ unknowns that are the PF coil currents.
In this example, there are $N_F=16$ Fourier harmonics per field line and $N_{\text{PF}}=5$ PF coils, resulting in 202 equations and 205 unknowns. That is, the system is underdetermined by construction, leaving ample freedom to satisfy the trace requirements.

\Cref{fig:cond1}(A-E) show a sequence of Poincare sections over the course of the continuation. \Cref{fig:cond1}(A, H) captures the initial state of the tracked fixed point, the purple circle with trace $T = 1.9$; the fixed point is elliptic, centered between two hyperbolic X-points. As the trace of the tracked fixed point approaches $T = 2$ (\Cref{fig:cond1}A-B), the island width shrinks. When the trace equals 2.0 exactly, as in \Cref{fig:cond1}(C), the bottom X-point and tracked O-point merge to become a rank-1 parabolic fixed point. \Cref{fig:cond1}(D) shows the tracked fixed point becomes hyperbolic, and a transient elliptic fixed point appears (black dot) through a transcritical bifurcation. 
Finally, in \Cref{fig:cond1}(D-E) the transient elliptic fixed point annihilates with the upper X-point collapsing to a single hyperbolic fixed point through a saddle-node bifurcation.
This rich behavior illustrates the control potential of this approach, and the complexity of stellarator divertor design. 

\Cref{fig:cond1}(F) shows that over the course of the continuation, the condition number (ratio of largest and smallest singular values) of the underdetermined Jacobian is well-behaved at the parabolic limit as well as in its neighborhood. In contrast, the condition number of the Jacobian associated with just the field line system, \cref{eq:continuation-rf-o}, is not, blowing up as the system approaches the parabolic fixed point. 
\Cref{fig:cond1}(G) prints the Z-coordinate of the different fixed points as they evolve and interact over the course of the continuation, showing a signal of the bifurcation.

\subsection{Snowflake: a rank-0 parabolic fixed point}
\label{sec:rank_0_example}

We now show an example of how formulation two, introduced in \Cref{sec:spectral_method}, can be used to compute a snowflake: a rank-0 parabolic fixed point with $\bm M = \bm I$. As in \Cref{sec:rank_1_example}, we perform a continuation on the trace of $\bm M$, $T$. The condition number of the Jacobian used in Newton's method remains well conditioned as the snowflake appears, $T=2$. Similar to \Cref{sec:rank_1_example}, we use fixed modular coils to produce a background magnetic field, while letting the currents in auxiliary PF coils vary to maintain existence of the fixed point.

\begin{figure}[tbh!]
    \centering
    \includegraphics[width=\textwidth]{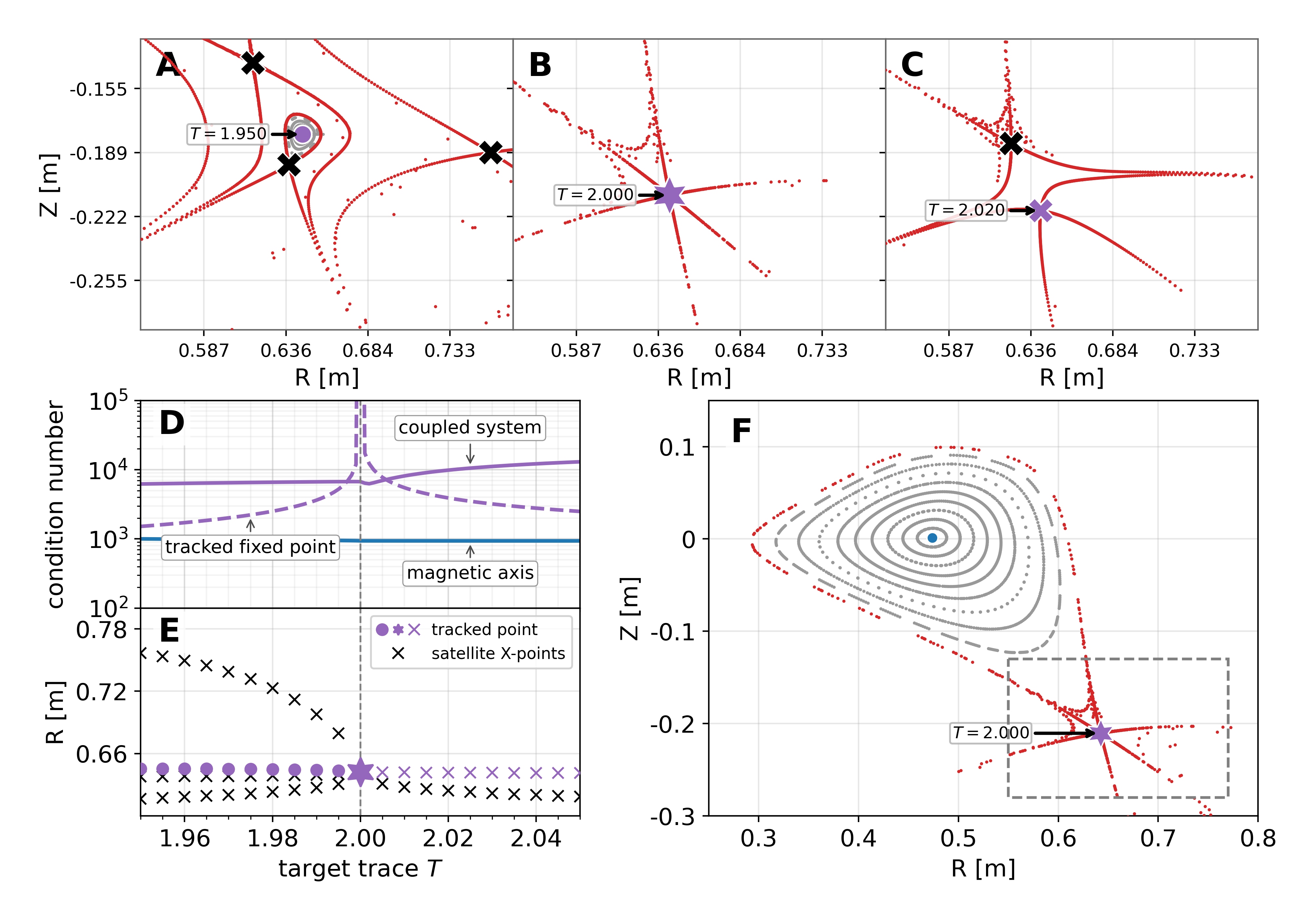}
    \caption{(A-C) Poincar\'e sections of the boxed region in (F) when the target trace value, $T$, of the tracked fixed point is varied about the bifurcation point. (F) An expanded view of the Poincar\'e section shown in (B), where the blue dot labels the magnetic axis. (D) Condition numbers of the Jacobian of \cref{eq:continuation2-rf-a} associated with the magnetic axis (blue), of \cref{eq:continuation2-rf-o} associated with the tracked fixed point (purple dashed), and of the full system \cref{eq:continuation2} (purple solid) as the target trace $T$ of the tracked fixed point is varied; the coupled system has a bounded condition number.
    (E) The $R(0)$ coordinates of the fixed points illustrating the bifurcation.
    Circle, cross, star markers correspond respectively to O, X, and parabolic fixed points.
    }
    \label{fig:cond2}
\end{figure}

Snowflakes in tokamaks are typically defined as configurations for which both the poloidal magnetic field and its first derivative vanishes \cite{ryutov_snowflake_2015}, i.e. $\mathbf{B}_\text{pol} = \nabla \mathbf{B}_\text{pol} = 0$. This is typically obtained in practice by pushing two X-points ``on top of each other" (that is, the snowflake could be considered the limit of taking two nearby X-points (and zero O-points) and making their separation vanish). 
For stellarators, a necessary condition to observe a snowflake is $\bm M=\bm I$.
The topological index of such a fixed point is $\sm 2$ (i.e. the winding number calculation taken around the snowflake would be $\sm 2$. From a distance, the snowflake has the same topological characteristic as two X-points, or three X-points and an O-point. As is observed in \cref{fig:cond2}, as $T$ increases or decreases from $T=2$, the snowflake will ``unfold" into these topologically equivalent states.

To observe the unfolding of a snowflake, we solve the following problem at each iteration of a continuation,
\begin{subequations}\label{eq:continuation2}
\begin{align}
    \mathbf r_f(\mathbf x, \mathbf I_{\text{PF}}) &= 0,\label{eq:continuation2-rf-a}\\
    \mathbf r_f(\mathbf t, \mathbf I_{\text{PF}}) &= 0,\label{eq:continuation2-rf-o}\\
    \Trace\!\left(M(\mathbf x, \mathbf I_{\text{PF}})\right) &= 1.6,\label{eq:continuation2-tr-a} \\
   \bm  M(\mathbf t, \mathbf I_{\text{PF}}) &= \mathbf m(T),\label{eq:continuation2-tr-o}
\end{align}
\end{subequations}
where the target tangent map is chosen to satisfy
$$
\mathbf m(T) = \begin{cases}
    \begin{pmatrix}
        T/2 & \sm \sqrt{4-T^2}/2 \\
         \sqrt{4-T^2}/2 & T/2
    \end{pmatrix} \text{ if } T < 2, \\ \\
    \begin{pmatrix}
        T/2 + \sqrt{T^2/4 - 1} & 0 \\
        0 & T/2 - \sqrt{T^2/4 - 1}
    \end{pmatrix} \text{ if } T \geq 2.
\end{cases}
$$
When $T=2$ holds precisely, $\bm M$ is the identity, and the fixed point corresponds to a snowflake. To make it simple to scan over values of $T$, we adopt the above parametrization of $\mathbf m$, though more general choices are possible when $T \neq 2$.

\Cref{fig:cond2}(A-C) shows Poincare sections over the course of the continuation. The continuation begins at $T=1.95$ in \Cref{fig:cond2}(A) where the elliptic fixed point is surrounded by three orbiting X-points. The snowflake appears as $T$ is increased to $T=2$ in \Cref{fig:cond2}(B), before splitting into two hyperbolic fixed points as $T$ is increased above 2. \Cref{fig:cond2}(F) shows an expanded view of the cross section with the snowflake.
In \Cref{fig:cond2}(D), we illustrate again how the standard discretization of \cref{eq:fieldline_2} is ill-suited to finding parabolic fixed points. Without augmenting the system with additional degrees of freedom and constraints, the Jacobian is ill conditioned as $T\to 2$. The Jacobian of the coupled system \cref{eq:continuation2}, on the other hand, is well-conditioned in any regime.

\subsection{A chain of O-points}
\begin{figure}
    \centering
    \includegraphics[width=\linewidth]{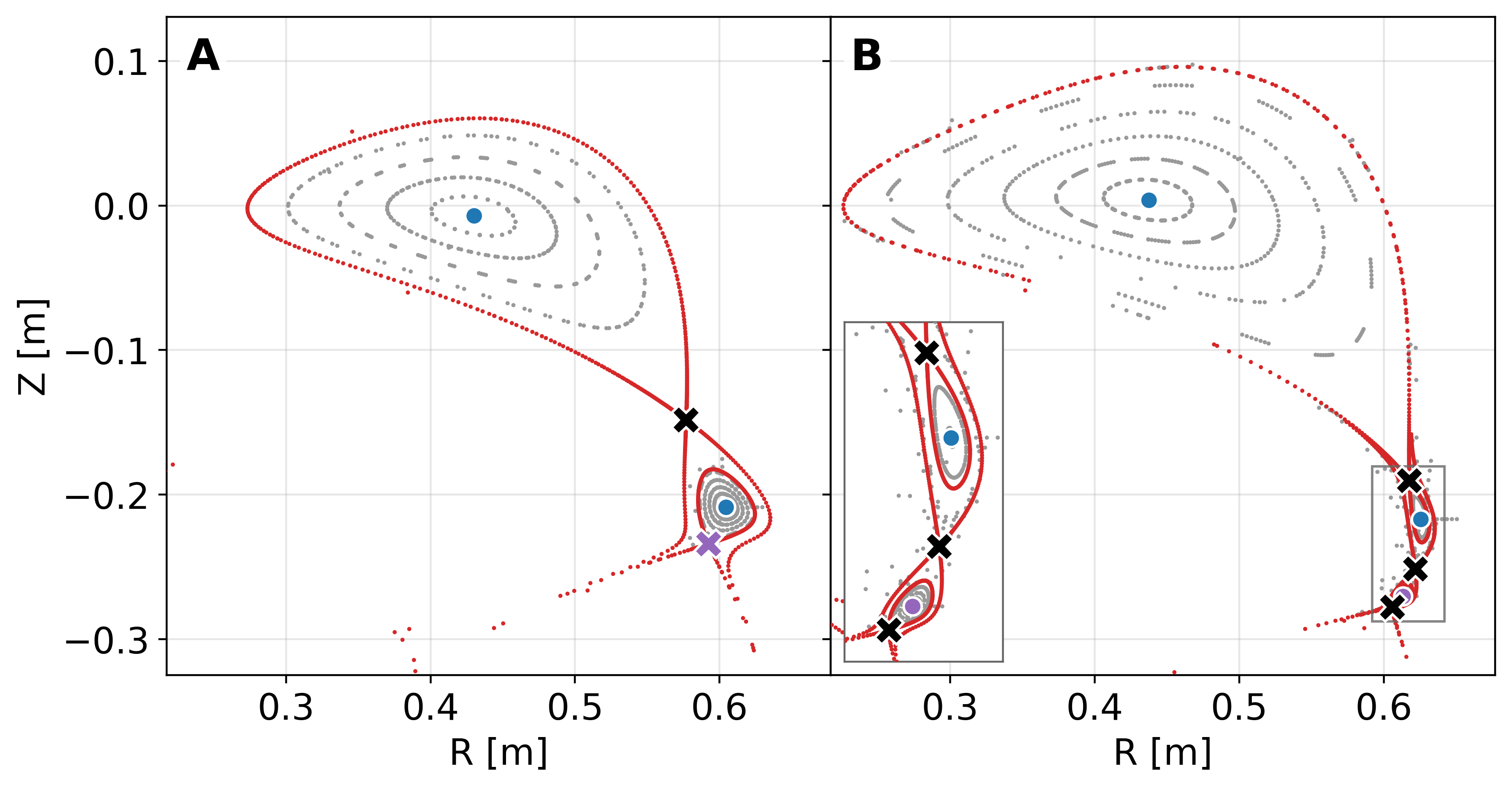}
    \caption{The tracked fixed point (purple) in panel A is converted from an X-point to an O-point in panel B, using a continuation. The traces of the tangent map for the remaining three fixed points in panel A are controlled to ensure they do not disappear. Converting the X-point into an O-point creates a divertor chain in panel B.}
    \label{fig:chain}
\end{figure}

As a final exercise of these new numerical methods, we convert the bottom X-point in \cref{fig:chain}A into an O-point, yielding a chain of O-points (\cref{fig:chain}B). 
To convert a $T > 2$ hyperbolic fixed point to a $T<2$ elliptic fixed point, the trace must pass through the $T=2$ parabolic limit. To do so,  the continuation problem is augmented by enforcing a trace constraint for each periodic fixed point \cref{fig:chain}A (the two elliptic points, and X-points). 
Specifically, we solve
\begin{subequations}
\begin{align} \label{eq:continuation3}
    \mathbf r_f(\mathbf x_i, \mathbf I_{\text{PF}}) &= 0,\\
    \mathbf r_f(\mathbf t, \mathbf I_{\text{PF}}) &= 0,\\
    \Trace\!\left(\bm M(\mathbf x_i, \mathbf I_{\text{PF}})\right) &= T_i,\\
    \Trace\!\left(\bm M(\mathbf t,  \mathbf I_{\text{PF}})\right) &= T,
\end{align} 
\end{subequations}
where $i=1,2,3$ indexes the three fixed points $\mathbf x_i$ whose traces are held at prescribed values $T_i$, while the bottom fixed point $\mathbf t$ is continued by varying the target trace $T$ from 1.95 to 2.05, so that it transitions from hyperbolic to parabolic snowflake, then elliptic.  System \eqref{eq:continuation3} constrains three field line  and trace conditions, resulting in 404 equations. 
It also has 410 unknowns corresponding to the field line geometries and $N_{\text{PF}}=10$ PF coil currents.
Again, there is ample freedom in this underdetermined system to satisfy the requested traces as the tracked fixed point's trace varies from 2.4 to 1.90.

\section{Vacuum vessel designs with simple signed distance functions}
\label{sec:vessel}

Realistic stellarator designs have a vacuum vessel that neither intersects with the coils nor the divertor fixed points. To automate finding such a vessel, the vessel geometry can be treated as a degree of freedom during the optimization, and varied to satisfy the compatibility constraints. To efficiently compute distances from the fixed points to the vessel, we introduce a family of vacuum vessel geometries for which signed distances from points to the vessel have closed-form expressions, or only require solving a one-dimensional root finding problem. The signed distance functions (SDFs) are differentiable and can be used inside objective functions or constraints in an optimization of the design.

A signed distance function, $\varrho(\mathbf p; \mathbf v)$, is the minimum Euclidean distance from a point $\mathbf p = (x,y,z)$ to the vessel $\Omega(\mathbf v)$, signed by whether $\mathbf p$ is inside or outside the volume. Explicitly,
\begin{equation}\label{eq:sdf}
\varrho(\mathbf p; \mathbf v) = \text{sgn}(\mathbf p; \Omega(\mathbf v)) \min_{\mathbf{p}^*\in \Omega(\mathbf v)} \|\mathbf p - \mathbf p^*\|,
\end{equation}
where $\mathbf v=(v_1, v_2, \ldots)$ are degrees of freedom parameterizing the vessel.
Throughout, a semicolon separates the spatial coordinates from the vessel's defining parameters, $\mathbf v$.
Points on the vessel are represented implicitly as the zero-level set of $\varrho$, i.e. $\{\mathbf p \, |\, \varrho(\mathbf p; \mathbf v)=0\}$. 
The vessel geometries used in this work have SDFs that are quick to compute and simple to implement. Because the SDF gives the distance to a parametrized geometry, the same machinery can also be used to parameterize ports and ensure clearance for port access \cite{baillod2025enhancing}.

\begin{figure}[tbh!]
    \centering
    \includegraphics[width=\linewidth]{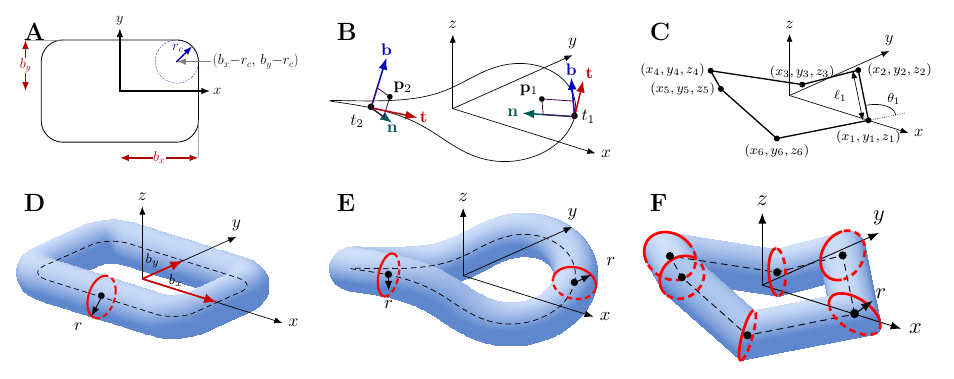}
    \caption{The three vacuum vessel families. Panels A and D show the pill pipe vessel, B and E the non-planar vessel, and C and F the piecewise-cylinder vessel realized as a loop of six mitred cylinders. A, B and C show the defining geometry (cross-section parameters for A, the centerline for B and C). D, E and F show the resulting three-dimensional vessels. The tube radius $r$ is indicated in the bottom row and the miter welds are highlighted in red in F.
    }
    \label{fig:pillpipe}
\end{figure}

A whole zoo of useful SDFs can be found in \cite{Quilez2013DistFunctions}.  
One family of surfaces that enables a quick SDF calculation is ``canal surfaces" \cite{PETERNELL_1997}. 
Such surfaces can be viewed as unions of spheres of radius $r(t)$ swept along a three-dimensional curve $\mathbf c(t)$. 
By observing that the distance from $\mathbf p$ to a sphere is $\|\mathbf p - \mathbf c(t)\| - r(t)$, it follows that the signed distance function at $\mathbf p$ is the solution to the minimization problem:
\begin{align}
    \varrho_{\text{canal}}(\mathbf p; \mathbf c, \mathbf r) = \min_{t \in [0, 1)} \|\mathbf p - \mathbf c(t)\| - r(t).
    \label{eq:canal_sdf}
\end{align}
A few useful vessel shapes with closed form SDFs can be expressed as canal surfaces with simple choices of $c(t), r(t)$.
The first canal surface we examine is called the pill pipe.
The pill pipe is the envelope of spheres with constant radius $r$ swept along a rounded rectangle of dimensions $2b_x \times 2b_y$ and corner radius $r_c$. These geometric parameters are illustrated in \Cref{fig:pillpipe}(A/D). 
The SDF for the pill pipe has a closed-form expression that is independent of $t$,
\begin{equation}
\begin{aligned}
\resizebox{0.9\linewidth}{!}{$\displaystyle
    \varrho(x, y, z; r,  b_x, b_y, r_c) = \sqrt{\left(\big\lVert \big(\max(a_x,0),\, \max(a_y,0)\big) \big\rVert_2 + \min\!\big(\max(a_x,a_y),\, 0\big) - r_c \right)^2 + z^2} - r
$},\\
a_x = |x| - (b_x - r_c), \qquad a_y = |y| - (b_y - r_c).
\end{aligned}
\end{equation}
The SDF remains valid when the side lengths of the rectangle and corner radius of the centerline are larger than the radius of the ball:
\begin{equation}\label{eq:validity_pill}
\begin{aligned}
    b_x - r_c, b_y-r_c, r_c&> 0, \\
    r &< r_c.
\end{aligned}
\end{equation}
A special case of the pill pipe when $c(t)$ is a circle of radius $R$ ($R = b_x=b_y=r_c$) is the simple toroidal vacuum vessel,
\begin{align}
    \varrho_{\text{torus}}(x, y, z; R, r) = \sqrt{(\sqrt{x^2 + y^2}-R)^2 + z^2}\, - r.
\end{align}

The next canal surface defines a vessel with a non-planar centerline $\mathbf c(t)$ swept by spheres with nonuniform radii $r(t)$, shown in \cref{fig:pillpipe}(B/E).
The formula for the radius is a standard Fourier series, and the centerline uses the same stellarator symmetric, $\nfp$-periodic parametrization as the periodic field lines \cref{eq:curvxyzfouriersymmetries1} and \cref{eq:curvxyzfouriersymmetries2}.
The SDF takes the general form of a one-dimensional minimization problem \cref{eq:canal_sdf}, which can be solved robustly using a Newton's method guarded by the bisection algorithm. 
Constraints on $c(t), r(t)$ are required to prevent ill-behaved geometries. First, we force that centerline to have uniform arc length, i.e. $\|\mathbf{c}'(t)\|$ is constant, by requesting $\mathrm{Var}_t\left[\|\mathbf{c}'(t)\| \right]=0$.  Next, we enforce geometric conditions to prevent non-differentiable creases in the vessel: for each $t\in[0,t_p)$,  we must have $r(t)> 0$, $r(t) < (1 - (r(t)^2)''/2)/\kappa(t)$ and $|r'(t)| < \|\mathbf{c}'(t)\|$ \cite{MaekawaTakashi1998Aaao,XuZhiqiang2006Aaap,PETERNELL_1997}, where $\kappa(t)$ is the curvature of the centerline. In the case of the pill pipe vessel, these generic conditions simplify to \eqref{eq:validity_pill}. While these conditions prevent local self-intersections and creases, they do not prevent the vessel from global self-intersections, i.e. portions of the vessel that are distant in the parameter $t$, but close geometrically \cite{MaekawaTakashi1998Aaao}.

The final vessel shapes that we consider are formed by a sequence of $n_{\mathrm{seg}}$ cylinders with radius $r$ that are glued together with the proper miter angle, shown in \cref{fig:pillpipe}(C/F). We call these ``piecewise cylindrical" (PC) vessels. The centerline of the cylinders is defined by a periodic piecewise linear interpolant through anchor points, where the centerline inherits the field period and stellarator symmetry of the device. Due to the sharp corners at the joints of the cylinders, this vessel class is not an instance of canal surfaces. 
The SDF for PC vessels requires solving a sequence of 1D minimizations, one per cylinder,
\begin{equation}\label{eq:sdf_pc}
\varrho_{\mathrm{pc}}(\mathbf p; \mathbf v) = \text{sgn}(\mathbf p; \mathbf \Omega(\mathbf v))\,\min_{\substack{\mathbf p^*\in \partial \Omega_k(\mathbf v)\\ k=1,\ldots,n_{\text{seg}}}} \|\mathbf p - \mathbf p^*\|,
\end{equation}
where $\partial \Omega_k$ is the boundary of $k$th cylinder, and the degrees of freedom are the vertices of the polyline and radius, $\mathbf v = (x_1, y_1, z_1, \hdots, r)$. \Cref{fig:pillpipe}(C) depicts the geometric degrees of freedom and the turn angles $\theta_k$. \Cref{eq:sdf_pc} computes the distance from $\mathbf p$ to the 
$n_{\mathrm{seg}}$ mitered cylinders. This is equivalent to computing the minimum of the distance to the $k$th infinite cylinder, and the two elliptic miter curves at the ends of the cylinder. Calculating the distance to an infinite cylinder has a simple closed form, but the distance to elliptical miters requires solving a nonlinear scalar equation which is described thoroughly in \cite{Eberly_DistancePointEllipse}.

\begin{figure}
    \centering
    \includegraphics[width=\linewidth]{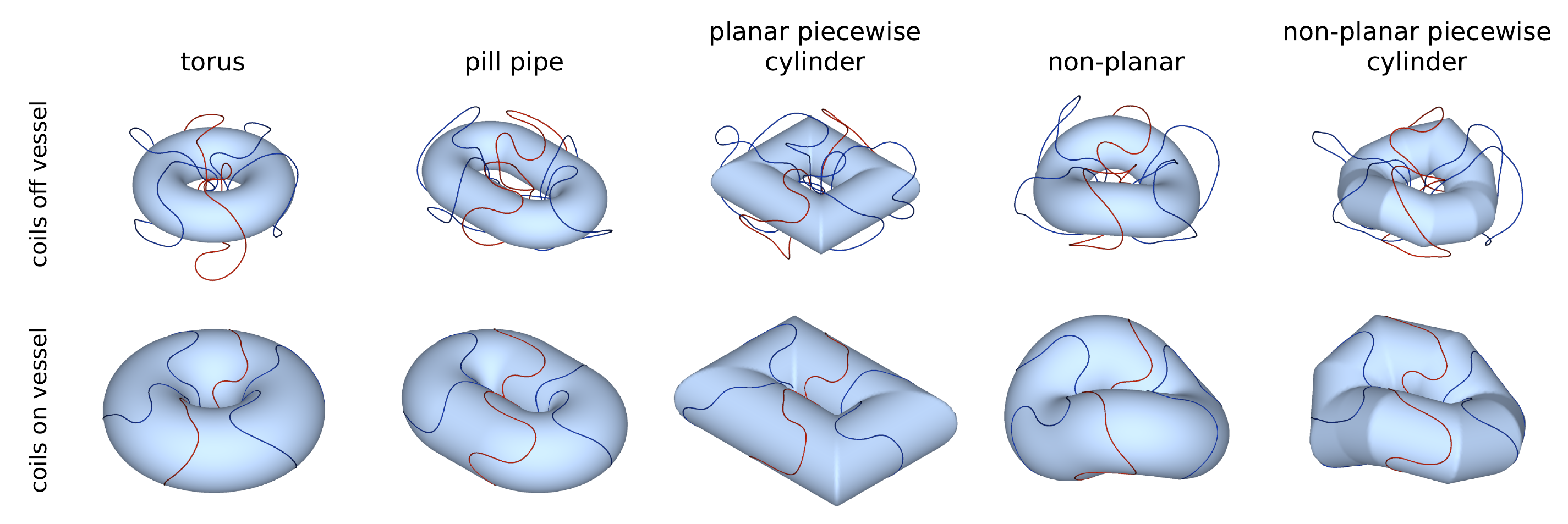}
    \caption{Overview of the vacuum vessel geometries supported by our signed-distance-function approach, shown for real optimized devices. Columns, from left to right: torus, pill pipe, planar piecewise cylinder, non-planar, and non-planar piecewise cylinder. The top row shows configurations with the coils placed off the vessel, the bottom row with the coils on the vessel. The two simultaneously optimized unique modular coil filaments are shown in red and blue}
    \label{fig:vessel_overview}
\end{figure}

We consider the PC vessel valid when the cylinders form a non-self-intersecting pipe. For a vessel defined by anchor points $\mathbf x_k = (x_k, y_k, z_k)$ and radius $r$, we require that $r > 0$, the length of each cylinder be non-negative, $\ell_k = \| \mathbf x_k - \mathbf x_{k+1}\| > 0$, each turn angle $\theta_k = \arccos\!\Big( (\mathbf x_k - \mathbf x_{k-1})\cdot(\mathbf x_{k+1} - \mathbf x_k)/(\ell_{k-1}\,\ell_k) \Big) < \theta_{\max} < 180^{\circ}$, and the two (elliptic) miters on each cylinder to not overlap, i.e. $\ell_k > r[\tan(\theta_k/2) + \tan(\theta_{k+1}/2)]$.  

In the special case of a PC vessel made up of $n_{\mathrm{seg}}=4$ perpendicular cylinders with centerline that lies in the XY-plane (shown in \Cref{fig:divertor_zoo}(first column)), we can find a level-set function with a closed form expression. We use the term level-set function to highlight that the function is not an exact signed distance function; the zero-level set ($\varrho_{\mathrm{pc}}(\mathbf p; \mathbf v)=0$) corresponds to the true vessel geometry, but in general, no longer provides a true distance to the vessel. In this case, the level set equation is,
\begin{align}\label{eq:approx_sdf}
    \varrho_{\mathrm{pc}}(x, y; b_x, b_y) =  \sqrt{\max\!\big(|x|-b_x,\ |y|-b_y\big)^2 + z^2} -r.
\end{align}

In \cref{sec:optimization}, we show how these SDFs can be included in the optimization to design vacuum vessels that are compatible with the fixed points and coils. \Cref{fig:divertor_zoo} showcases five vessel geometries found using stellarator optimization over vessel geometries.
All the SDFs mentioned here are differentiable with respect to vessel shape parameters. 
There are some instances when the nearest point on the vessel to $\mathbf p$ is not unique, for example, when $\mathbf p$ lies on the centerline, $\bm c(t)$, of a canal vessel. When this occurs, the derivative of the SDF with respect to $\mathbf p$ is not defined. While this could be problematic for a gradient-based stellarator optimization routine, it is not an issue as long as the optimizer does not query the spatial gradient through these problematic areas. 
In practice, it did not prevent our algorithm from finding useful devices.
An overview of the various STAR\_Lite class devices with compatible vacuum vessels are shown in \cref{fig:vessel_overview}.

\section{Divertor and vacuum vessel direct optimization} 
\label{sec:optimization}
In this section, we incorporate the differentiable methods for field line control (\Cref{sec:ODE}) and vacuum vessel design (\Cref{sec:vessel}) into a single framework to jointly design coils, divertor fixed points, and a vacuum vessel for a stellarator. The optimization must be initialized from a configuration where a divertor fixed point exists.  Though not used here, we expect that a $\bm B \times d\bm l$ minimization \cite{FLOM2026115846} could be used to generate such an initialization if needed in future.

Our optimization formulation treats the vacuum vessel geometry as a degree of freedom with parameters $\mathbf v$, as well as any coil currents and geometries, specified by the vector $\mathbf q$. As discussed in \Cref{sec:vessel}, we choose a vacuum vessel parameterization that admits a signed distance function, $\varrho$. The fixed point position $\bm{\Gamma}_{\text{divertor}}(t \, ; \mathbf q)$ can be computed from the magnetic field using the techniques from \Cref{sec:spectral_method}. Additional degrees of freedom for shaping the magnetic field must be introduced in order to stabilize the computation; in all numerical experiments we introduce auxiliary PF coils with variable radii and currents. 

When optimizing the divertor placement and vacuum vessel geometry, it is relevant to constrain the pairwise distances between the vessel, fixed-point, and coils. To that end, we include the following pairwise distances constraints in the joint optimization,
\begin{subequations}\label{eq:codesign}
\renewcommand{\theequation}{\theparentequation.\arabic{equation}}
\begin{align}
  & d_{\text{coil-vessel}} \le \varrho\big(\bm{\Gamma}_{\text{coil}}(t \, ; \, \mathbf{q});\, \mathbf{v}\big) \le D_{\text{coil-vessel}}
    && \text{[coil-vessel distance]} \label{eq:coil_vessel_constraint} \\
  & \varrho\big(\bm{\Gamma}_{\text{divertor}}(t\, ;\, \mathbf{q});\, \mathbf{v}\big) \le d_{\text{divertor-vessel}}
    && \text{[divertor-vessel distance]} \label{eq:divertor_vessel_constraint}  \\
  & \varrho\big(\bm{\Gamma}_{\text{surface}}(\varphi, \theta \, ;\, \mathbf{q}) \, ;\, \mathbf{v}\big) \le d_{\text{surface-vessel}}
    && \text{[opt. surface-vessel distance]} \label{eq:surface_vessel_constraint} \\
  & \mathrm{Var}_t[\varrho\big(\bm{\Gamma}_{\text{divertor}}(t \, ;\,\mathbf{q}) \, ;\, \mathbf{v}\big)] = 0
    && \text{[const. divertor-vessel distance]} \label{eq:const_divertor_vessel_constraint} \\
  & \mathrm{Var}_t\big[\,\mathbf{e}_z\cdot\bm{\Gamma}_{\text{divertor}}(t \, ;\, \mathbf{q})\,\big] = 0
    && \text{[const. divertor $Z$-coordinate]} \label{eq:const_divertor_Z_constraint}
\end{align}
\end{subequations}
$\bm{\Gamma}_{\text{surface}}, \bm{\Gamma}_{\text{coil}}$ are vectors representing the position of points on any flux surface, and coils, respectively.
\Cref{eq:coil_vessel_constraint} constrains the pairwise distance between each coil and the vacuum vessel to be at least $d_{\text{coil-vessel}}$ and at most $D_{\text{coil-vessel}}$. While the lower bound is used to satisfy packing constraints, the upper bound can be used to keep the coils close to the vessel, or even put the coils on the vessel, as shown in \Cref{sec:example_configs}.
\Cref{eq:divertor_vessel_constraint} ensures that the fixed point does not intersect with the vessel; since $\varrho$ is an SDF and the fixed point is on the interior of the vessel, $d_{\text{divertor-vessel}}$ is negative. 
\Cref{eq:surface_vessel_constraint} constrains the distance between a flux surface of the magnetic field and the vessel -- $d_{\text{surface-vessel}}$ is also negative.
The two variance constraints \cref{eq:const_divertor_vessel_constraint,eq:const_divertor_Z_constraint} additionally hold the divertor fixed point at a constant distance from the vessel and a constant vertical position as it winds toroidally. The variance is computed statistically at quadrature points on the divertor fixed point. While \cref{eq:const_divertor_vessel_constraint,eq:const_divertor_Z_constraint} are not strictly necessary, including them can simplify the process of designing divertor plates by keeping the divertor strikelines at fixed positions in a cross section. 

When seeking a configuration with a parabolic fixed point, the optimization may be seeded with a configuration with an elliptic or hyperbolic fixed point. If the initial configuration has an elliptic or hyperbolic fixed point and $\Trace(\bm M)$ is far from $2$, then the additional degrees of freedom reserved for fixed point computation, may not have enough flexibility to convert the fixed point to parabolic. To remedy this, first we drive the configuration close to a parabolic fixed points to get an improved initial guess. Then the auxiliary degrees of freedom will have enough flexibility to ensure the parabolic fixed point can be found, so the spectral methods from \cref{sec:spectral_method} can be applied using the exact $\Trace(\bm M) = 2$, or $\bm M = \bm I$ constraint. To find an improved initial configuration, we augment \eqref{eq:codesign} with a constraint on the return map $\bm M$. We introduce the constraint,
\begin{equation} \label{eq:rank1}
    |\Trace( \bm M(\mathbf{c}, \mathbf{I}, \mathbf{x}(\mathbf{c}, \mathbf{I}))) - 2| \leq 0.1
\end{equation}
when seeking rank-1 parabolic fixed points and 
\begin{equation}\label{eq:rank0}
    \| \bm M(\mathbf{c}, \mathbf{I}, \mathbf{x}(\mathbf{c}, \mathbf{I})) - \bm I \|_{\infty} \leq 0.1 
\end{equation}
when seeking rank-0 parabolic fixed points.  
The solutions to \eqref{eq:codesign} augmented with exactly one of \eqref{eq:rank1} or \eqref{eq:rank0} correspond to stellarators whose divertors are close to parabolic in the desired subclass.  
We use them as initial guesses in a final optimization restricted to the space of stellarators with perfectly parabolic divertors.

\subsection{Application to designing STAR\_Lite-class stellarators}
\label{sec:optimization_problem_for_starlite}
We now extend the optimization formulation beyond the basic pairwise distance constraints used for vacuum vessel and fixed point design, \cref{eq:codesign}, to find STAR\_Lite-class stellarators. The goal of the optimization is to find STAR\_Lite scale stellarators with a quasisymmetric region of nested flux surfaces, a certain rotational transform, a magnetic well, a certain aspect ratio, a fixed point for a divertor, and a compatible vacuum vessel. Some configurations resulting from this optimization are shown in \cref{sec:example_configs}. To ensure the existence of a parabolic fixed point,  when seeking one, one or more additional circular PF coils with variable radii, vertical position, and current are introduced.

The geometric constraints in \cref{eq:codesign} are coupled with a standard stellarator coil design problem,
\begin{subequations}\label{eq:codesign2}
\renewcommand{\theequation}{\theparentequation.\arabic{equation}}
\begin{align}
\min_{\mathbf{q},\,\mathbf{v}} \quad
  & f_{\mathrm{QA}}\big(\mathbf{q}\big)
    && \notag \\
\text{subject to}\quad
  & \iota\big(\mathbf{q}\big) = \iota_s^{*}
    && \text{[opt. surface rotational transform]} \label{eq:iota_surface_constraint} \\
  & \iota\big(\mathbf{q}\big) = \iota_a^{*}
    && \text{[on-axis rotational transform]} \label{eq:iota_axis_constraint} \\
  & A\big(\mathbf{q}\big) = A^{*}
    && \text{[aspect ratio of opt. surface]} \label{eq:aspect_constraint} \\
  & \int_{0}^{1} \big\| \mathbf{B}\big(\bm{\Gamma}_{\text{axis}}(t; \mathbf{q})\big) \big\|\, \mathrm{d}t = B^{*}
    && \text{[mean field strength]} \label{eq:mean_modB_constraint} \\
  & \operatorname{Var}_{t}\!\Big[\, \big\| \mathbf{B}\big(\bm{\Gamma}_{\text{axis}}(t \, ;\, \mathbf{q})\big) \big\| \,\Big] = 0
    && \text{[on-axis field variance]} \label{eq:axis_modB_variance_constraint} \\
  & W\big(\Psi \,; \,\mathbf{q}\big) \le W^{*}
    && \text{[magnetic well]} \label{eq:well_constraint}  \\
  & |\mathbf{I}(\mathbf{q})| \le \mathbf{I}^{*}
    && \text{[coil current bound]} \label{eq:current_constraint}
\end{align}
\end{subequations}
The objective $f_{\mathrm{QA}}(\mathbf q)$ measures deviation from quasisymmetry on a target optimization surface $\bm \Gamma_{\mathrm{surface}}(\varphi, \theta; \mathbf{q})$, parameterized in Boozer coordinates, see \cref{sec:qs_error}.
This flux surface is computed from the magnetic field using the ``Boozer surface method": a PDE with residual $\rb_s(\sbold, \mathbf q)=0$ is solved, defining the mapping $\sbold(\mathbf q)$ and the rotational transform on the surface. See \cite{giuliani_single-stage_2022, giuliani_direct_2022} for a full treatment of the approach. The magnetic axis, $\bm{\Gamma}_{\text{axis}}(t;\mathbf q)$, is computed from the magnetic field using the spectral method from \cref{sec:spectral_method}.

\Cref{eq:iota_surface_constraint} fixes the rotational transform, $\iota$, on the target flux surface to $\iota_s^{*}$, and \cref{eq:iota_axis_constraint} fixes $\iota$ on the magnetic axis via \eqref{eq:iota_a} to $\iota_a^{*}$. Together, these two constraints control the shear. \Cref{eq:aspect_constraint} fixes the aspect ratio of the optimization surface to a target $A^{*}$. $A^{*}$ is set to that of STAR\_Lite design A \cite{harrer_star_lite_2026} ($A^*  =6.66$), though the last closed flux surface might be lower aspect ratio.

The two field strength conditions, \cref{eq:mean_modB_constraint} and \cref{eq:axis_modB_variance_constraint}, set the mean field strength on-axis to target $B^{*}=0.0875 \mathrm{T}$, and enforce on-axis quasisymmetry, respectively.
\Cref{eq:well_constraint} imposes a magnetic well, which is favorable for MHD interchange stability (see \cref{sec:well} for details on how the well term is calculated).

\Cref{eq:current_constraint} bounds each coil current by a coil-specific maximum collected in $\mathbf{I}^{*}$, where the absolute value and inequality act elementwise. Manufacturing constraints set the upper bound to $60~\mathrm{kA} \cdot \mathrm{turns}$ for the modular coils and we impose $5~\mathrm{kA}  \cdot \mathrm{turns}$ for the PF coils when they are present.
We also impose additional geometric constraints on the vacuum vessel geometry, ensuring that it remains valid and non-self-intersecting; see \cref{sec:vessel} for these additional details.
We impose additional engineering constraints adopted by the STAR\_Lite project \cite{harrer_star_lite_2026} on uniform incremental arclength of the coils, coil-to-coil distance, maximum and mean-squared coil curvature, coil length, major radius.  The complete problem, including these constraints and their bounds, is stated in \cref{sec:full_problem}.

Optimization is initialized from STAR\_Lite design A, the baseline quasi-axisymmetric coil set of the STAR\_Lite experiment \cite{harrer_star_lite_2026} whose divertor is of X-point type, and solve \eqref{eq:codesign} using a penalty method \cite{nocedal_numerical_2006}. Sensitivities with respect to the coils and currents ($\mathbf q$) in all cases can be obtained using automatic differentiation.  Note that we do not differentiate through the Newton solve.  Rather, we form the adjoint system using JAX and compute discretely exact gradients using vector-Jacobian products.
The ODE and PDE constraints are imposed exactly, while a penalty method attempts to satisfy engineering and physics constraints to 0.1\% accuracy when the target is nonzero, and 0.1 absolute error when the target is zero.

\section{Example configurations} \label{sec:example_configs}
In this section, we solve the optimization problem introduced in \cref{sec:optimization_problem_for_starlite} to design STAR\_Lite class stellarators with novel divertor structures. 
The devices differ in the divertor type, divertor location, vessel shape, and relative location of the coils: on the vessel or off the vessel, and whether there are PF coils or not.

\begin{figure}
    \centering
    \includegraphics[width=0.98\linewidth]{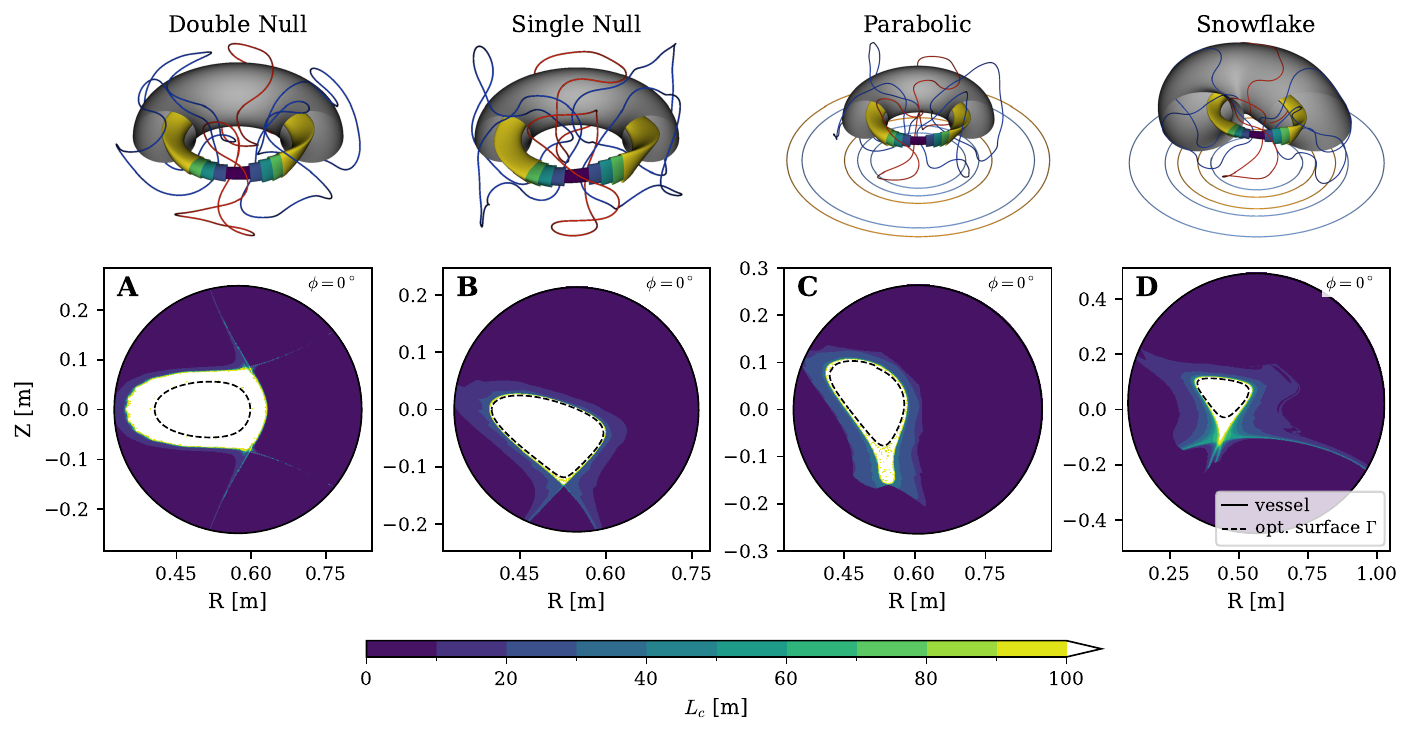}
     \caption{(Top row) Four optimized stellarators with diverse diverting fixed point topologies.
      The modular coils are displayed in blue and red, and the poloidal field coils are shown in orange and light blue.
      The vacuum vessel is shown revealing the nested flux surfaces within.
      (Bottom row) Connection length $L_c$ on the $\phi =0$ cross section for the devices in the top row. The circular vessel cross section (solid line) also serves as the divertor target in this exercise, and the black dashed line is the cross section of the optimization surface $\Gamma$ on which quasisymmetry is targeted.
     }
    \label{fig:example_montage}
\end{figure}

Since a full characterization of the tool's capabilities, as well as the physical implications of the resulting structures for a reactor, is far beyond the scope of this paper, we present only a small subset of the possible structures, along with a preliminary physical analysis.
To this end, we introduce four new configurations in \cref{fig:example_montage}. 
The first two are relatively simple double-null (DN) and single-null (SN) X-point divertors, \cref{fig:example_montage}(A-B). The remaining two examples showcase more exotic varieties, with panel (C) showing a parabolic single-null divertor and panel (D) a snowflake.
All four are two-field-period STAR\_Lite class devices with six modular coils (blue and red in the top row of \cref{fig:example_montage}), enclosed by a vacuum vessel with circular cross sections.
The parabolic and snowflake configurations additionally employ poloidal field (PF) coils, shown in orange and light blue, to shape and polish the diverting fixed point. Note that these five circular auxiliary coils are non-uniformly spaced. 
While the double-null configuration retains stellarator symmetry, the single-null configurations necessarily break it: a diverting fixed point that sits only below (or only above) the plasma is incompatible with the up-down mirror symmetry that stellarator symmetry imposes on the $\phi=0$ cross section. As discussed below, this has consequences for how much of the device must be inspected when analyzing the scrape-off layer.

\subsection{Connection length and strike lines}

\begin{figure}
    \centering
    \includegraphics[width=0.98\linewidth]{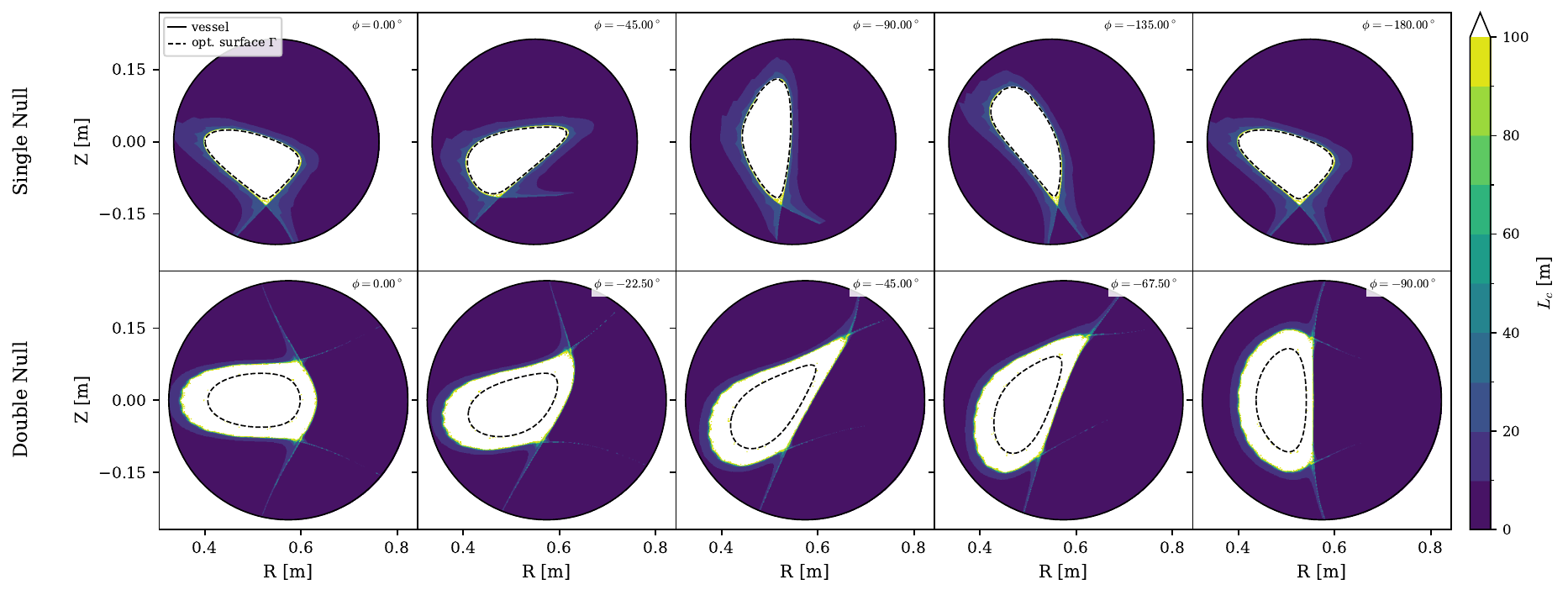}
     \caption{Connection length $L_c$ on a sequence of toroidal cross sections for the single-null (top row) and double-null (bottom row) configurations of \cref{fig:example_montage}. Since the single null breaks stellarator symmetry, the cross sections span the full field period $\phi \in [0^\circ, -180^\circ]$; for the stellarator symmetric double null the half field period $\phi \in [0^\circ, -90^\circ]$ determines the entire device. The solid line is the vessel cross section and the dashed line the optimization surface $\Gamma$. In both configurations the confined core and the divertor legs persist smoothly across all toroidal angles.}
    \label{fig:lc_phi_montage}
\end{figure}

\begin{figure}[h]
    \centering
    \includegraphics[width=\linewidth]{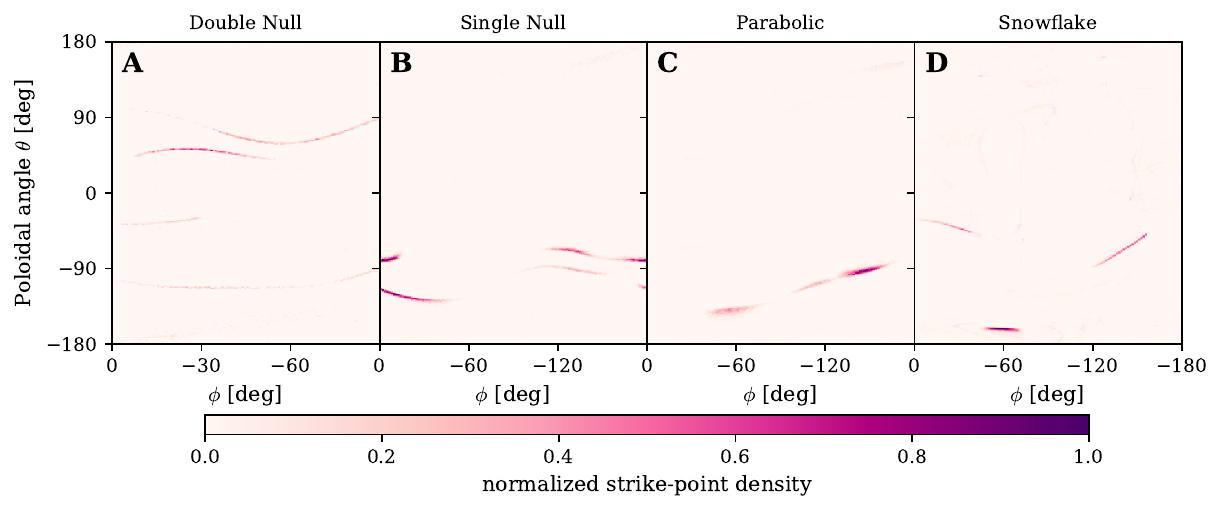}
     \caption{Strike lines for the four devices from \cref{fig:example_montage}. Zero degrees of the poloidal angle represents the outboard midplane with the positive direction following the vessel in the counter clockwise direction.}
    \label{fig:strike_density_montage}
\end{figure}

The connection length plots (second row of \cref{fig:example_montage} and \cref{fig:lc_phi_montage}) and strike-point plots (\cref{fig:strike_density_montage}) were created using the FLARE library \cite{Frerichs_2024_FLARE}, which allows fast field line tracing and analysis. In both cases, the magnetic field is evaluated on a precomputed grid with $256 \times 256 \times 128$ points in $(R, Z, \phi)$ per field period. The connection length maps are computed by launching a field line from every node of a $512 \times 512$ grid in $(R,Z)$ on each cross section shown. Each field line is traced at most 50\,m in both the forward and backward directions or until it hits the vessel, and the two lengths are summed to give the connection length $L_c$ represented by the color bar gradient. At the major radius $R_0 = 0.5$\,m, the 50\,m cap corresponds to roughly 15 toroidal transits per direction, well beyond the connection lengths of the divertor legs, so it truncates only confined and near-confined field lines that never reach the wall.
The connection length maps in the second row of \cref{fig:example_montage} share a common structure: a region of long connection length ($L_c > 100$\,m, saturated in the color bar) marks the confined core, which encloses the optimization surface $\Gamma$ in all four configurations. Across the separatrix, $L_c$ drops sharply, and the divertor legs are visible as narrow bands of intermediate connection length that guide the open field lines from the fixed point to the vessel wall. The distinct fixed-point topologies are clearly reflected in the leg structure: the double null exhibits the expected up-down symmetric pairs of legs, the single null and the parabolic configuration channel the exhaust into legs below the plasma, and the snowflake displays the characteristic additional legs emanating from its rank-0 parabolic fixed point.

A single toroidal cross section is, however, not sufficient to assess divertor behavior in a stellarator. \Cref{fig:lc_phi_montage} therefore shows the connection length on a sequence of cross sections along the device for the single-null (top row) and double-null (bottom row) configurations. Here the broken stellarator symmetry of the single null becomes important: since cross sections at $\phi$ and $-\phi$ are no longer mirror images of one another, the full field period $\phi \in [0^\circ, -180^\circ]$ must be covered, whereas for the stellarator symmetric double null the half field period $\phi \in [0^\circ, -90^\circ]$ suffices, with the remaining cross sections following by symmetry. Both configurations behave well over the entire device: the core region of long connection length remains intact at every toroidal angle, the sharp gradient of $L_c$ across the separatrix persists, and the divertor legs sweep smoothly and continuously along the vessel wall rather than appearing or vanishing abruptly.

The strike plots in \cref{fig:strike_density_montage} are generated using the optimized vessels as preliminary strike targets. Field lines are launched from a flux surface just outside the separatrix on a grid of $180 \times 720$ points ($129\,600$ field lines) and traced in both directions for up to 100\,m, with a small artificial cross-field diffusion coefficient of $10^{-5}\,\mathrm{m^2/m}$ and a maximum integration step of $0.0175$\,rad. Their intersections with the vessel are binned onto a $240 \times 120$ grid in poloidal and toroidal angle, spanning one field period for the non-stellarator-symmetric configurations and one half period for the stellarator symmetric double null, whose remaining half follows by symmetry. The intersection count is normalized by the maximum count per cell for each configuration individually and shown as a color bar; this displays the shape of each strike pattern but not its absolute magnitude, so we additionally compare the configurations on a common scale below.

In all four configurations the strike points organize into narrow bands that are well localized in both the poloidal and toroidal angle, rather than being smeared over the vessel: the wetted area is set by the divertor topology, with the double null distributing the exhaust over four up-down symmetric bands while the single-null configurations concentrate it below the midplane ($\theta \approx -90^\circ$). 
Strike line maps indicate where physical divertor targets would need to be placed on each device, and confirm that the diverted field lines strike the vessel in an ordered, topology-determined pattern rather than diffusely.

For a comparison across the configurations, we consider the smallest wall area that receives half of all field-line strikes, $A_{50\%}$, normalized by the respective vessel surface area, since the four vessels differ substantially in size. While $A_{50\%}$ is not a standard figure of merit, it is a quantile-based analogue of the wetted-area measures used in tokamak heat-load studies, chosen here because it is insensitive to the histogram bin resolution and to the number of field lines traced. By this measure the snowflake spreads the bulk of its exhaust most effectively: half of its strikes are distributed over $1.5\%$ of the vessel surface, compared to $0.7\%$ for the double null and $0.4$--$0.5\%$ for the single-null and parabolic configurations, a relative improvement by a factor of $2$--$3.5$. The snowflake divertor is effective at spreading heat, though we caution that it might not be attributed to the divertor topology alone: the snowflake is also the most chaotic of the four edge magnetic fields, which by itself broadens the strike pattern. Disentangling these contributions, and establishing whether the observed ordering holds systematically, will require a statistical analysis over many configurations, which we plan to carry out on the larger database of devices mentioned in \cref{sec:conclusions}.

\subsection{MHD Stability Proxy}
The configurations of \cref{fig:example_montage} are vacuum fields and are therefore ideal-MHD stable by construction: with $p' \equiv 0$ there is no pressure drive. The relevant question is instead whether the novel edge topologies come at the price of MHD stability once a plasma is present: the strong shaping required to produce a parabolic or snowflake fixed point could, in principle, degrade the interchange stability of the core. To address this question at the operating point of the planned experiment, we evaluate two complementary ideal-MHD criteria: the Mercier interchange criterion, as computed by VMEC \cite{hirshman_whitson_1983}, and the ballooning growth rate, computed with COBRAVMEC \cite{sanchez_cobra_2000}, both evaluated on a fixed-boundary equilibrium based on the vacuum field with a low volume-averaged $\beta = 0.01\%$, matching the parameters expected for STAR\_Lite design A \cite{harrer_star_lite_2026}. Both are linear, ideal, local criteria that are necessary conditions for stability rather than predictions of the achievable $\beta$. 
Moreover, at such low $\beta$ the pressure drive is weak, so passing these criteria confirms the configurations are stable at the operating point of the experiment; we defer analysis at reactor-relevant $\beta$   to the finite-$\beta$ extension of the framework discussed in the conclusions.

The Mercier criterion tests for local interchange instabilities, in which neighboring flux tubes swap radial positions without appreciably bending the magnetic field \cite{goldston_1995_mercier, Kamenetskii_1972_mercier}. It decomposes into four terms, $D_\mathrm{Merc} = D_\mathrm{Shear} + D_\mathrm{Well} + D_\mathrm{Curr} + D_\mathrm{Geod}$, with the configuration Mercier-stable wherever the sum is positive: the magnetic shear term $D_\mathrm{Shear} \geq 0$ is stabilizing, the geodesic curvature term $D_\mathrm{Geod} \leq 0$ is destabilizing, and $D_\mathrm{Well}$ is stabilizing in the presence of a magnetic well. $D_\mathrm{Curr}$ reflects the contribution of the net toroidal current and can take either sign; since our configurations are current-free, this term is negligible here. Since $D_\mathrm{Well}$ can be evaluated directly on the vacuum field, we are able to optimize for it explicitly, as shown through Equation (\ref{eq:well_constraint}).

The ballooning criterion complements Mercier's criterion by testing for instabilities that are localized along a field line rather than at a point, and which are driven by pressure gradients coupling to unfavorable local curvature. Unlike Mercier's criterion, ballooning modes can grow even where the flux-surface-averaged curvature is favorable, making the two criteria complementary rather than redundant. We report $\gamma_\mathrm{max}$, the maximum over the sampled flux surfaces and field lines of the signed ballooning growth rate computed by COBRAVMEC \cite{sanchez_cobra_2000}, evaluated on $11$ flux surfaces spanning $s = 0.05$ to $0.95$ with $4 \times 4$ initial angular positions per surface; $\gamma_\mathrm{max} < 0$ indicates stability. An independent calculation of the ballooning eigenvalue with a one-dimensional finite-element solver on the same equilibria confirms stability on all analyzed surfaces.

\begin{figure}[h]
    \centering
    \includegraphics[width=0.98\linewidth]{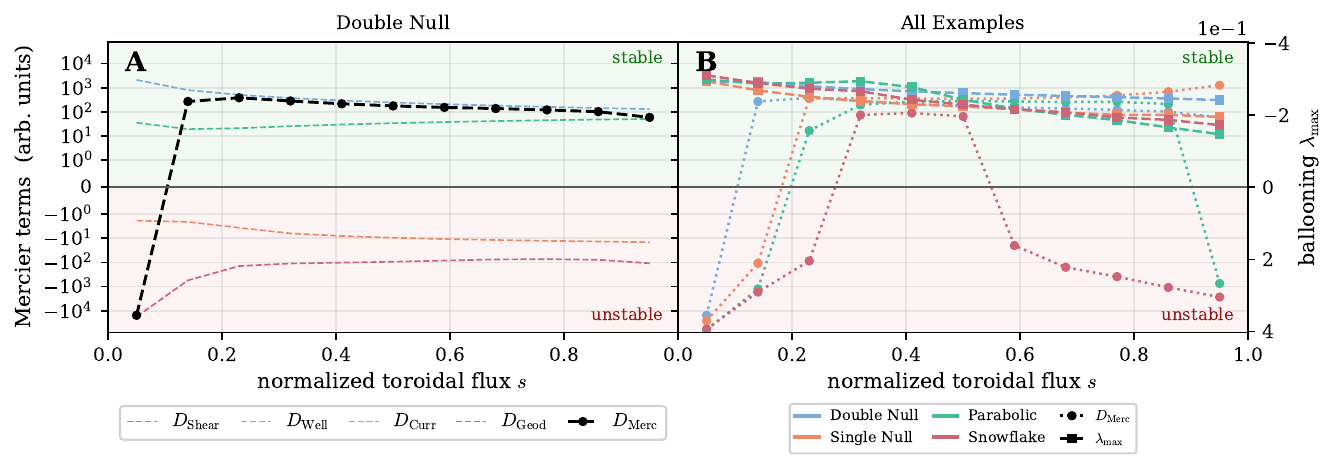}
    \caption{
    Ideal MHD stability analysis via the Mercier and ballooning criteria. Panel A shows the individual Mercier terms ($D_\mathrm{Shear}$, $D_\mathrm{Well}$, $D_\mathrm{Curr}$, $D_\mathrm{Geod}$) and their sum $D_\mathrm{Merc}$ as a function of the normalized toroidal flux $s = \Psi/\Psi_b$, where $\Psi_b$ is the flux at the plasma boundary, for the double-null configuration; the configuration is Mercier-stable wherever $D_\mathrm{Merc} > 0$. Panel B compares $D_\mathrm{Merc}$ (left axis, circles) and the maximum ballooning growth rate $\gamma_\mathrm{max}$ computed with COBRAVMEC \cite{sanchez_cobra_2000} (right axis, squares) across all four configurations. Note that the Mercier terms are given in arbitrary VMEC units; only their sign and relative composition are physically meaningful, not their absolute magnitude.}
    \label{fig:mhd_montage}
\end{figure}

\Cref{fig:mhd_montage} summarizes the analysis. Panel A shows the Mercier criterion split into its constituent terms $D_\mathrm{Shear}$, $D_\mathrm{Well}$, $D_\mathrm{Curr}$, and $D_\mathrm{Geod}$ for the double-null configuration, illustrating the balance described above: away from the magnetic axis, the destabilizing geodesic contribution is overcome by the combination of shear and the optimized magnetic well. Panel B shows the ballooning growth rate $\gamma_\mathrm{max}$ and $D_\mathrm{Merc}$ for all four configurations. With respect to ballooning modes, all four devices, including the parabolic and snowflake configurations, are stable ($\gamma_\mathrm{max} < 0$) on every flux surface analyzed. The Mercier criterion is more restrictive. The double-null, single-null, and parabolic configurations are Mercier stable over the bulk of the plasma volume, with an unstable region near the magnetic axis ($s \lesssim 0.1$ for the double null, $s \lesssim 0.2$ for the single null and the parabolic configuration) and, for the parabolic configuration, a return to instability on the outermost surfaces. We note that the innermost flux surfaces, where all four configurations return $D_\mathrm{Merc} < 0$, are also where VMEC's evaluation of the Mercier criterion is known to be least reliable, so the near-axis instability should be interpreted with caution. The snowflake configuration is Mercier stable only in the intermediate region $0.3 \lesssim s \lesssim 0.5$. We conclude that the exotic divertor topologies are compatible with ballooning stability at STAR\_Lite parameters, and that Mercier stability over most of the plasma volume is attainable at least up to the parabolic divertor class. For the snowflake, interchange stability is the binding constraint: notably, this configuration was optimized with the magnetic well constraint active ($V'' \leq -100$), which for this topology is evidently not sufficient to obtain Mercier stability beyond the intermediate region. 
A systematic study of how Mercier stability trades against divertor topology and quasisymmetry, across the larger database of configurations mentioned in the conclusions, is left to future work.

Finally, we verify in \cref{sec:well} that the magnetic well constraint \cref{eq:well_constraint} indeed has the intended effect on these stability measures, by comparing two devices with identical design targets that differ only in whether the well is targeted (\cref{fig:well_opt_impact}).

\section{Conclusions}
\label{sec:conclusions}
This work presents, for the first time, a set of algorithms for the simultaneous optimization of stellarator equilibria, arbitrary divertor topology and vacuum vessel. This is achieved by the development of two new computational tools: a differentiable method for computing periodic field lines and the Jacobian of their return map, which is well-conditioned even in the degenerate limit of parabolic fixed points; and an efficient scheme for designing vessels which do not intersect the confined region, edge fixed points or coils, achieved by using signed distance functions. 

The algorithms presented here are generally applicable to stellarator design, but here we demonstrate their applicability and flexibility by designing next-generation candidates for the  STAR\_Lite university-scale stellarator experiment at Hampton University, Virginia \cite{harrer_star_lite_2026}. These stellarators are optimized for quasi-axisymmetric vacuum fields, buildable coils, geometrically simple vessels and novel divertor topologies. The examples demonstrate the versatility of the algorithms; in addition to single- and double-null-like configurations, we also show divertor structures never previously reported in stellarators; six-legged ``snowflake" divertors (which are targeted as periodic field lines of the parabolic ``flavor") and an ``O-point chain" which diverts plasma to the vessel wall but without (as in island divertors) encircling the confined plasma. The divertor performance of such topologies, when scaled to reactor scale, is currently unknown (and also depends on divertor plate and baffle geometry, which is not explored here), but this work presents an avenue by which divertor topology can be easily and efficiently manipulated to explore such advanced concepts. The proposed algorithms are robust enough to generate a data set of STAR\_Lite class stellarators with advanced divertors, which will be analyzed in future work.

The promising results of the algorithms presented here motivate their application beyond the design of STAR\_Lite configurations. 
Extending the optimization to stellarators with finite plasma $\beta$, would require the plasma-generated magnetic field to be included in the calculation of the edge fixed points. 
Including the plasma-generated field could be done, for example, using a virtual casing principle such as calculated by BIEST \cite{malhotra2020efficient} or EXTENDER \cite{drevlak2005pies}, or by using a topology-agnostic equilibrium code such as SPEC \cite{loizu2016verification, hudson2020free}. 
Additional target functions could be included in the optimization to parametrize divertor performance. These might include flux expansion and upstream-to-target connection lengths as is used to quantify tokamak divertor performance \cite{soukhanovskii_developing_2018, soukhanovskii2022first, gorno2023power, harrison2024benefits}, low-fidelity stellarator edge models such as two-point models \cite{feng2011comparison, maaziz2026investigating} or reduced heat transport models \cite{frerichs2026firefly, feng2022review, Frerichs_2024_FLARE, kharwandikar2025power}, or more expansive high-fidelity models such as EMC3-EIRENE \cite{feng2014recent} or BOUT++-based tools \cite{shanahan2019fluid, shanahan2024global, tork2025bout, bold2026numerical}. Such calculations usually rely on the plasma-facing component geometry as well as the magnetic geometry, but a number of relatively straightforward algorithms for automated divertor plate/baffle construction already exist which could be used as a starting point \cite{davies2024semi, davies_squid_2025, liu2025universal, veksler2026stellarator, frerichs2026firefly}.

\section{Data availability}

The scripts that generate all data presented in this paper will be made publicly available in a repository on Zenodo.

\section{Acknowledgments}
Supported by DOE-RENEW (DE-SC0025698, SN, GW), DOE-FAIR (DE-SC0024443, CL), the Simons Foundation (1167550, CL), and the Friedrich Schiedel Foundation for Energy Technology (RW).
This work has been carried out within the framework of the EUROfusion Consortium, partially funded by the European Union via the Euratom Research and Training Programme (Grant Agreement No 101052200 — EUROfusion). Views and opinions expressed are however those of the author(s) only and do not necessarily reflect those of the European Union or the European Commission. Neither the European Union nor the European Commission can be held responsible for them. We thank the Flatiron Institute's Scientific Computing Core.
We also thank Nikita Nikulsin for sharing his ballooning stability code.
The authors acknowledge the use of various AI systems for assistance with manuscript formulation and plotting code. All AI-assisted content was reviewed, edited, and verified by the authors, who take full responsibility for the integrity of the final work. 

\appendix

\section{Full optimization problem}
\label[appendix]{sec:full_problem}
For completeness, we state here the full optimization problem \eqref{eq:codesign}, augmented with the engineering constraints that were omitted from \cref{sec:optimization} for brevity. Re-using the numbering of \cref{sec:optimization}, the problem reads
\begin{align}
\min_{\mathbf{q},\,\mathbf{v}} \quad
  & f_{\mathrm{QA}}\big(\mathbf{q},\, \mathbf{s}(\mathbf{q})\big)
    && \notag \\
\text{subject to}\quad
  & \iota\big(\mathbf{s}(\mathbf{q})\big) = \iota_s^{*}
    && \text{[opt. surface rotational transform]} \tag{\ref{eq:iota_surface_constraint}} \\
  & \iota\big(\mathbf{a}(\mathbf{q})\big) = \iota_a^{*}
    && \text{[on-axis rotational transform]} \tag{\ref{eq:iota_axis_constraint}} \\
  & A\big(\mathbf{s}(\mathbf{q})\big) = A^{*}
    && \text{[aspect ratio]} \tag{\ref{eq:aspect_constraint}} \\
  & \int_{0}^{1} \big\| \mathbf{B}\big(\bm{\Gamma}_{\text{axis}}(t; \mathbf{a}(\mathbf{q}))\big) \big\|\, \mathrm{d}t = B^{*}
    && \text{[mean field strength]} \tag{\ref{eq:mean_modB_constraint}} \\
  & \operatorname{Var}_{t}\!\Big[\, \big\| \mathbf{B}\big(\bm{\Gamma}_{\text{axis}}(t;\mathbf{a}(\mathbf{q}))\big) \big\| \,\Big] = 0
    && \text{[on-axis field variance]} \tag{\ref{eq:axis_modB_variance_constraint}} \\
  & W\big(\Psi; \mathbf{a}(\mathbf{q}),\, \mathbf{s}(\mathbf{q}),\, \mathbf{q}\big) \le W^{*}
    && \text{[magnetic well]} \tag{\ref{eq:well_constraint}} \\
  & |\mathbf{I}(\mathbf{q})| \le \mathbf{I}^{*}
    && \text{[coil current bound]} \tag{\ref{eq:current_constraint}} \\
  & d_{\text{coil-vessel}} \le \varrho\big(\bm{\Gamma}_{\text{coil}}(t;\mathbf{q});\, \mathbf{v}\big) \le D_{\text{coil-vessel}}
    && \text{[coil-vessel distance]} \tag{\ref{eq:coil_vessel_constraint}} \\
  & \varrho\big(\bm{\Gamma}_{\text{divertor}}(t;\mathbf{x}(\mathbf{q}));\, \mathbf{v}\big) \le d_{\text{divertor-vessel}}
    && \text{[divertor-vessel distance]} \tag{\ref{eq:divertor_vessel_constraint}} \\
  & \varrho\big(\bm{\Gamma}_{\text{surface}}(\varphi, \theta;\mathbf{s}(\mathbf{q}));\, \mathbf{v}\big) \le d_{\text{surface-vessel}}
    && \text{[opt. surface-vessel distance]} \tag{\ref{eq:surface_vessel_constraint}} \\
  & \mathrm{Var}_t[\varrho\big(\bm{\Gamma}_{\text{divertor}}(t;\mathbf{x}(\mathbf{q}));\, \mathbf{v}\big)] = 0
    && \text{[const. divertor-vessel distance]} \tag{\ref{eq:const_divertor_vessel_constraint}} \\
  & \mathrm{Var}_t\big[\,\mathbf{e}_z\cdot\bm{\Gamma}_{\text{divertor}}(t;\mathbf{x}(\mathbf{q}))\,\big] = 0
    && \text{[const. divertor $Z$-coordinate]} \tag{\ref{eq:const_divertor_Z_constraint}} \\
  & \operatorname{Var}_{t}\!\big[\, \| \bm{\Gamma}_{\text{coil},i}'(t;\mathbf{q}) \| \,\big] = 0
    && \text{[uniform incremental arclength]} \tag{\ref{eq:codesign}.13} \label{eq:arclength_constraint} \\
  & \big\| \bm{\Gamma}_{\text{coil},i}(\mathbf{q}) - \bm{\Gamma}_{\text{coil},j}(\mathbf{q}) \big\|_{2} \ge d_{\text{cc}}
    && \text{[coil-to-coil distance]} \tag{\ref{eq:codesign}.14} \label{eq:coil_coil_constraint} \\
  & L_i(\mathbf{q}) \le L_{\max}
    && \text{[coil length]} \tag{\ref{eq:codesign}.15} \label{eq:length_constraint} \\
  & \max_{t}\, \kappa_i(t;\mathbf{q}) \le \kappa_{\max}
    && \text{[maximum curvature]} \tag{\ref{eq:codesign}.16} \label{eq:curvature_constraint} \\
  & \frac{1}{L_i(\mathbf{q})}\int_0^1 \kappa_i(t;\mathbf{q})^2\, \| \bm{\Gamma}_{\text{coil},i}'(t;\mathbf{q}) \|\, \mathrm{d}t \le \kappa_{\mathrm{msc}}
    && \text{[mean-squared curvature]} \tag{\ref{eq:codesign}.17} \label{eq:msc_constraint} \\
  & R\big(\mathbf{s}(\mathbf{q})\big) = R_0
    && \text{[major radius]} \tag{\ref{eq:codesign}.18} \label{eq:major_radius_constraint}
\end{align}
for each coil $i$ and each pair of distinct coils $i \neq j$. Here $\bm{\Gamma}_{\text{coil},i}(t;\mathbf{q})$ denotes the $i$-th coil, $\kappa_i(t;\mathbf{q})$ its curvature, and $L_i(\mathbf{q}) = \int_0^1 \| \bm{\Gamma}_{\text{coil},i}'(t;\mathbf{q}) \|\, \mathrm{d}t$ its length.

Constraints \eqref{eq:iota_surface_constraint}--\eqref{eq:const_divertor_Z_constraint} are described in \cref{sec:optimization}. The remaining relations are the additional engineering constraints, whose bounds follow the values used for design A of the STAR\_Lite project \cite{harrer_star_lite_2026}. \Cref{eq:arclength_constraint} parameterizes each coil by uniform incremental arclength. \Cref{eq:coil_coil_constraint} keeps distinct coils at least $d_{\text{cc}} = 0.15~\mathrm{m}$ apart. \Cref{eq:length_constraint} limits each coil to a length $L_{\max} = 3~\mathrm{m}$. \Cref{eq:curvature_constraint,eq:msc_constraint} bound the maximum and mean-squared curvatures by $\kappa_{\max} = 10.4~\mathrm{m}^{-1}$ and $\kappa_{\mathrm{msc}} = 20.0~\mathrm{m}^{-2}$, respectively. Finally, \cref{eq:major_radius_constraint} fixes the major radius of the optimization surface to $R_0 = 0.5~\mathrm{m}$.

Separately, the vacuum vessel geometry is constrained to remain valid and non-self-intersecting, with the constraints depending on the vessel family (\cref{sec:vessel}):
\begin{subequations}\label{eq:vessel_geometry_constraints}
\begin{align}
\left.
\begin{aligned}
&0 < r_c < b_x, \quad r_c < b_y \\
&0 < r < r_c, \quad r < r_c
\end{aligned}
\right\} &\quad \text{pill pipe vessel} \label{eq:vessel_pill} \\[4pt]
\left.
\begin{aligned}
&\operatorname{Var}_{t}\!\big[\, \|\mathbf{c}'(t;\mathbf v)\| \,\big] = 0 \\
&r(t;\mathbf v) > 0 \\
&r(t; \mathbf v) < \frac{1 - (r(t;\mathbf v)^2)''/2}{\kappa(t)} \\
&|r'(t;\mathbf v)| < \|\mathbf{c}'(t;\mathbf v)\|
\end{aligned}
\right\} &\quad \text{canal vessel} \label{eq:vessel_np} \\[4pt]
\left.
\begin{aligned}
&r > 0, \quad \ell_k > 0, \quad \theta_k < \theta_{\max} \\
&\ell_k > r\big[\tan(\theta_k/2) + \tan(\theta_{k+1}/2)\big]
\end{aligned}
\right\} &\quad \text{piecewise-cylinder vessel} \label{eq:vessel_pc}
\end{align}
\end{subequations}
where $\kappa(t; \mathbf v)$ is the curvature of the non-planar centerline $\mathbf{c}(t; \mathbf v)$, and, for the piecewise-cylinder vessel, $\ell_k$ and $\theta_k$ are the length of the $k$-th cylinder and the turn angle at the $k$-th joint, for $k = 1, \dots, n_{\text{seg}}$.

\section{Quasisymmetry objective}
\label[appendix]{sec:qs_error}

Suppose a flux surface with degrees of freedom $\sbold$ and is parameterized by Boozer angles $\theta\in[0,1]$ and $\varphi\in[0,1/n_{\text{fp}}]$.
The objective $f_{\mathrm{QA}}$ measures the average violation from quasisymmetry, where the individual quasisymmetry violation is
\begin{equation}\label{eq:qa1}
f_{\mathrm{QA}}(\cb, \Ib, \sbold) = \frac{\iint_S(B(\varphi, \theta;\cb, \Ib, \sbold) - B_{\text{QA}}(\theta;\cb, \Ib, \sbold) )^2  ~dS}{\iint_S B_{\text{QA}}(\theta;\cb, \Ib, \sbold)^2 ~dS},
\end{equation}
and
\begin{equation}\label{eq:qa2}
B_{\text{QA}}(\theta;\cb, \Ib, \sbold) = \frac{\int_0^{1/n_{\text{fp}}} B(\varphi,\theta; \cb, \Ib, \sbold) \|\mathbf n(\varphi, \theta ;\mathbf s) \|~d\varphi}{\int_0^{1/n_{\text{fp}}} \|\mathbf n(\varphi, \theta; \sbold)\|~d\varphi}
\end{equation}
is the closest quasisymmetric field strength, in a least squares sense, to the true one generated by the coils. Here the integrals are taken over the targeted optimization surface. Note that $B_{\text{QA}}$ depends only on $\theta$: for a quasi-axisymmetric field, the field strength on a flux surface is a function of $\theta$ alone in Boozer coordinates, so $B_{\text{QA}}(\theta)$ is the field strength of the closest quasi-axisymmetric field. See \cite{giuliani_direct_2022} for the full derivation.

\section{Controlling the magnetic well}
\label[appendix]{sec:well}
The magnetic well property
$$
W := \frac{d^2V}{d\Psi^2} < 0,
$$
is desirable to prevent the appearance of interchange modes, where $V$ is the volume of a flux surface and $\Psi$ is the toroidal flux \cite{Landreman_Jorge_2020}. This section shows how to compute the magnetic well from two or more surfaces parameterized in Boozer coordinates. Note that one of the surfaces can be the magnetic axis.

Given a set of flux surfaces with toroidal fluxes $\Psi_j$ for $j=1,\ldots, N$, the magnetic well can be computed by fitting an order-$2N+1$ Hermite polynomial interpolant to the volume, and differentiating it twice with respect to $\Psi$. The Hermite interpolant, $\hat{V}$, interpolates the volume $V_j$ and derivative of the volume $V'_j$, on each flux surface,
\begin{equation}
\begin{aligned}
\hat{V}(\Psi_j) &= V_j, \quad j=1,\ldots, N\\
\hat{V}'(\Psi_j) &= V'_j, \quad j=1,\ldots, N.
\end{aligned}
\end{equation}
The volume of a surface parameterized in Boozer coordinates can be computed through a boundary integral using divergence theorem. The derivative of the volume with respect to $\Psi$ can be computed by differentiating the volume integral,
\begin{equation} \label{eq:boozer_jac}
V(\Psi) = \int_0^{\Psi} \int_0^1 \int_0^1 \frac{G(\psi)+\iota I(\psi)}{B(\psi, \varphi, \theta)^2} ~d\psi ~d\varphi ~d\theta,
\end{equation}
which simplifies to,
\begin{equation}\label{eq:vprime}
V'(\Psi) = \int_0^1 \int_0^1 \frac{G}{B(\Psi, \varphi, \theta)^2} ~d\varphi ~d\theta,
\end{equation}
when the magnetic field is a curl-free $(I=0,  G=\text{const})$. The right-hand-side of \eqref{eq:vprime} can be evaluated directly on a Boozer surface, providing the derivative data to construct a Hermite polynomial for the volume.

In practice, the interpolation conditions are evaluated from data taken from one Boozer surface and the magnetic axis, producing a cubic hermite polynomial. Differentiating the model twice produces a linear model for the magnetic well. Note that the magnetic well is not unitless and is affected by the device's scale and field strength.

During optimization, such as in \Cref{sec:optimization}, the magnetic well can be controlled via a constraint of the form,
\begin{align}
W\big(\Psi; \mathbf{a}(\mathbf{q}),\, \mathbf{s}(\mathbf{q}),\, \mathbf{q}\big)=\hat{V}''(\Psi) \leq W^*
\end{align}
where $W^*$ is a target value of the magnetic well. The constraint can be upheld pointwise.

\begin{figure}
     \centering
        \includegraphics[width=\linewidth]{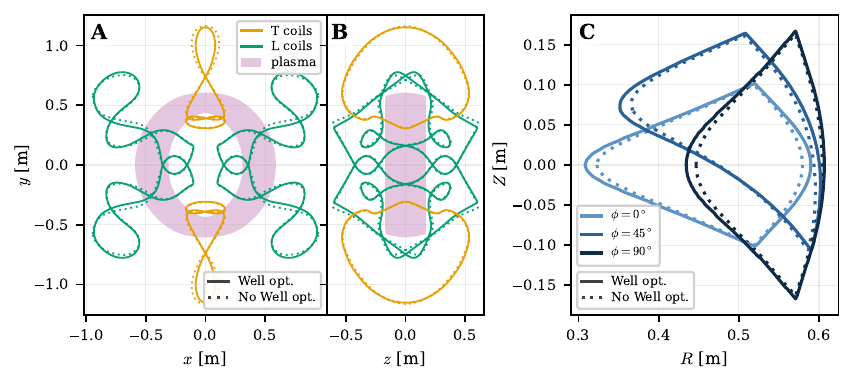}
    \includegraphics[width=\linewidth]{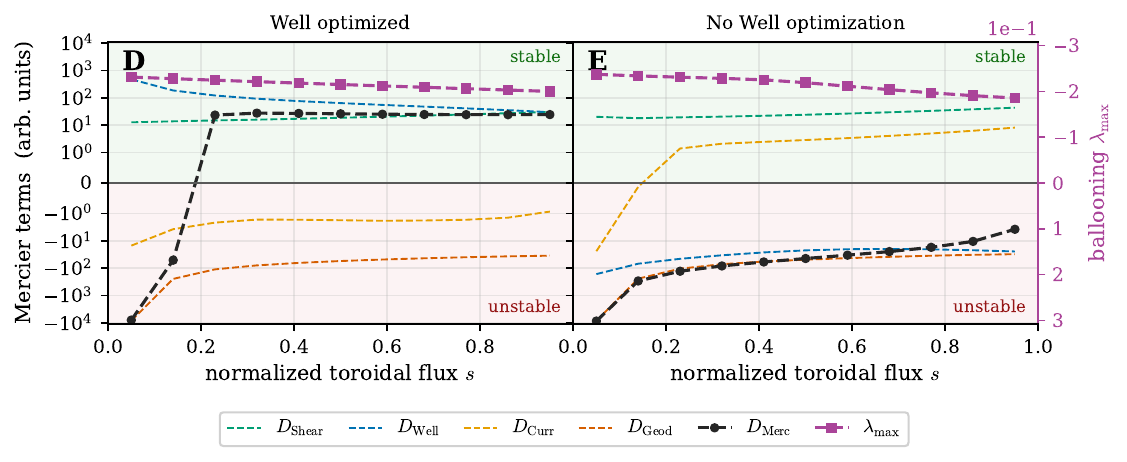}
         \caption{Impact of well optimization on a particular device. We compare two configurations with identical optimization goals, differing only in the well optimization: the solid lines correspond to a configuration with a well-term target of $V'' \leq -100$, while the dotted lines correspond to a configuration in which the well term was not targeted.  MHD stability changes when well optimization is utilized.}
     \label{fig:well_opt_impact}
\end{figure}

To confirm that the well constraint has the intended effect on the MHD stability proxies of \cref{sec:example_configs}, we compare two STAR\_Lite class devices with the same design targets, except that one targets magnetic well and the other does not (\cref{fig:well_opt_impact}). The trade-off between requesting a magnetic well and the achieved quality of quasisymmetry has already been studied in \cite{lp_precise}, where it was revealed that requesting well is typically possible, but comes at the detriment of the quasisymmetry; we echo this observation here. Note that visually, the modular coils and surface cross sections of the two configurations are quite close to one another. This hints that the magnetic well might be particularly sensitive to manufacturing errors, and a good candidate for risk neutral optimization.

\bibliographystyle{abbrv}
\bibliography{references}
\end{document}